\documentclass[twocolumn,twocolappendix]{aastex7}

\usepackage{CJK}
\usepackage[colorinlistoftodos]{todonotes}
\usepackage[export]{adjustbox}
\usepackage{bm}
\usepackage{amsmath}
\usepackage{amssymb}
\usepackage{graphicx} 
\usepackage{float} 
\usepackage{subfigure}
\usepackage{enumerate} 
\usepackage{booktabs}
\usepackage{listings}
\usepackage{makecell}
\usepackage{enumitem} 
\usepackage{placeins} 

\hypersetup{colorlinks=true, citecolor=blue, linkcolor=blue,
  urlcolor=magenta} \makeatletter

\newcommand{\Rmnum}[1]{\expandafter\@slowromancap
  \romannumeral #1@} \makeatother

\usepackage{aas_macros}
\usepackage{amsfonts}
\usepackage{float}
\usepackage{amsmath,amssymb}
\usepackage{natbib}    
\usepackage{hyperref} 
\usepackage{graphicx} 
\usepackage{color}
\usepackage{verbatim}
\usepackage{enumitem}
\usepackage{natbib}

\usepackage[linesnumbered,ruled]{algorithm2e}

\newcommand{\proptosim}{\mathrel{\vcenter{
 \offinterlineskip\halign{\hfil$##$\cr
 \propto\cr\noalign{\kern2pt}\sim\cr\noalign{\kern-2pt}}}}}
\renewcommand{\b}[1]{\boldsymbol{#1}}
\newcommand{\unit}[1]{{\rm #1}}

\newcommand{\response}[1]{{\color{blue} #1}}
\newcommand{\placeholdfigure}[2][]{%
  \begin{figure}[htbp]
    \centering
    \fbox{%
      \parbox{0.7\textwidth}{%
        \centering\vspace{2cm}%
        \textbf{[Placeholder: #2]}%
        \if\relax\detokenize{#1}\relax\else\\[0.3cm]\small #1\fi
        \vspace{2cm}%
      }%
    }
    \if\relax\detokenize{#1}\relax
      \caption{#2}
    \else
      \caption{#1}
    \fi
    \if\relax\detokenize{#2}\relax\else\label{#2}\fi
  \end{figure}
}
\newcommand{\placeholdtable}[2][]{%
  \begin{table}[htbp]
    \centering
    \fbox{%
      \parbox{0.5\textwidth}{%
        \centering\vspace{1.5cm}%
        \textbf{[Placeholder: #2]}%
        \if\relax\detokenize{#1}\relax\else\\[0.3cm]\small #1\fi
        \vspace{1.5cm}%
      }%
    }
    \if\relax\detokenize{#1}\relax
      \caption{#2}
    \else
      \caption{#1}
    \fi
    \if\relax\detokenize{#2}\relax\else\label{#2}\fi
  \end{table}
}

\newcommand{\au}{\mathrm{AU}}
\newcommand{\cm}{\unit{cm}}
\newcommand{\m}{\unit{m}}
\newcommand{\g}{\unit{g}}
\newcommand{\G}{\unit{G}}
\newcommand{\K}{\unit{K}}
\newcommand{\km}{\unit{km}}

\newcommand{\eV}{\unit{eV}}
\newcommand{\keV}{\unit{keV}}
\newcommand{\s}{\mathrm{s}}
\newcommand{\yr}{\mathrm{yr}}

\newcommand{\ang}{\ensuremath{\mathrm{\AA}}}
\newcommand{\dyn}{\mathrm{dyn}}

\newcommand{\lya}{\text{Ly}\ensuremath{\alpha~}}

\newcommand{\B}{\mathbf{B}}     
\newcommand{\E}{\mathbf{E}}     
\newcommand{\J}{\mathbf{J}}     
\renewcommand{\v}{\mathbf{v}}   

\renewcommand{\d}{\mathrm{d}}

\newcommand{\e}{\mathrm{e}}

\renewcommand{\ion}[2]
{{\rm#1}\;\textsc{\MakeLowercase{#2}}}
\newcommand{\p}{\mathrm{p}}     

\newcommand{\dust}{\mathrm{dust}}

\renewcommand{\O}{\mathrm{O}}     
\newcommand{\A}{\mathrm{A}}     
\renewcommand{\H}{\mathrm{H}}     
\renewcommand{\P}{\mathrm{P}}     
\newcommand*\chem[1]{\ensuremath{\mathrm{#1}}}
\newcommand{\pos}[1]{\ensuremath{\mathrm{#1}^+}}
\renewcommand{\neg}[1]{\ensuremath{\mathrm{#1}^-}}
\newcommand{\ext}[1]{\ensuremath{\mathrm{#1}^*}}

\newcommand{\code}[1]{\lstinline{#1}}
\newcommand{\kratos}{\code{Kratos}}

\newcommand{\figdir}{.} 

\renewcommand{\response}[1]{#1}

\begin{document}
\begin{CJK*}{UTF8}{gbsn}

\title{Modeling the Dynamics and Thermochemistry for the
    Outer Atmospheres of the \\ Ultra-hot Jupiter WASP-121b}

\author[0000-0002-6540-7042]{Lile Wang (王力乐)}
\affiliation{The Kavli Institute for Astronomy and
  Astrophysics, Peking University, Beijing 100871, China}
\affiliation{Department of Astronomy, School of Physics,
  Peking University, Beijing 100871, China}
\email{lilew@pku.edu.cn}

\author[0000-0003-2509-6558]{Yiren Lin (林伊人)}
\affiliation{The Kavli Institute for Astronomy and
  Astrophysics, Peking University, Beijing 100871, China}
\affiliation{Department of Astronomy, School of Physics,
  Peking University, Beijing 100871, China}
\email{lilew@pku.edu.cn}

\author[0000-0002-4361-8885]{Ji Wang (王吉)}
\affiliation{Department of Astronomy, The Ohio State
  University, 100 W 18th Ave, Columbus, OH 43210 USA}
\email{wang.12220@osu.edu}

\author[0000-0002-8958-0683]{Fei Dai (戴飞)}
\affiliation{Institute for Astronomy, University of Hawaii,
  2680 Woodlawn Drive, Honolulu, HI, 96822, USA} 
\email{fdai@hawaii.edu}

\correspondingauthor{Lile Wang}
\email{lilew@pku.edu.cn}

\begin{abstract}
  We present three-dimensional simulations of the ultra-hot
  Jupiter (UHJ) WASP-121b from the planetary surface to
  extended outflows, coupling hydrodynamics with consistent
  non-equilibrium thermochemistry, ray-tracing radiative
  transfer, and hydrodynamics using the GPU-accelerated
  \kratos{} framework. The fiducial model exhibits several
  atmospheric layers, including the lower atmospheres
  controlled by day-night circulation, and transonic
  photoevaporative outflows at higher altitudes shaped into
  two spiral arms by the stellar gravity and orbital motion
  effects. Different species could trace different regions:
  Fe probes rotation-dominated inner layers, Na maps dense
  spiral arms where recombination balances photoionization,
  and H$\alpha$ and He $10830~\ang$ features trace
  progressively more extended, ionized gas.  With spiral arm
  velocities reaching $\sim 40~\km~\s^{-1}$ projected along
  the line of sight, this morphology naturally reproduces
  the velocity pattern of observed high-velocity Na and
  H$\alpha$ absorption features without requiring
  significant super-rotation jet streams, although the
  absolute absorption amplitudes could carry uncertainties
  from stellar UV luminosity and trace elemental abundances.
  Parametric studies reveal complex dependencies on stellar
  irradiation: enhanced FUV intensifies outflows and extends
  spiral arms spatially and kinematically, while EUV and
  X-ray expands spiral structures into attenuated, ionized
  regions.  Stellar wind confinement compresses the dayside
  outflow and enhances metastable helium absorption. This
  work demonstrates that current and future transmission
  spectral observations that probe multiple species can
  provide important constraints on astrophysical
  environments of UHJs by comparing state-of-the-art
  simulations.
\end{abstract}

\section{Introduction}
\label{sec:intro}

Ultra-Hot Jupiters (UHJs) represent an extreme class of gas
giants with dayside equilibrium temperatures
$\gtrsim 2000~\K$, driven by intense stellar irradiation at
orbital distances $\lesssim 0.1~\au$ \citep{Parmentier_2018,
  Lothringer_2018}.  These highly irradiated planets
experience intense stellar fluxes that fundamentally shape
their atmospheric structure, causing significant atmospheric
inflation and lifting atmospheric layers to high altitudes
\citep{2018ApJ...855L..30A, Komacek_2018}.  Extreme thermal
conditions drive atmospheric chemistry into a regime where
molecular bonds are dissociated, or even atoms could be
ionized, resulting in atmospheres composed predominantly of
atomic and ionic species rather than intact molecules
\citep{Evans_2017,Parmentier_2018}.  This unique
thermochemical state, combined with their tidally locked
rotation that creates permanent dayside and nightside
hemispheres, distinguishes UHJs from their cooler Hot
Jupiter counterparts and makes them laboratories for
studying planetary atmospheres under the most extreme
conditions observable.

UHJs could serve as testbeds for atmospheric
characterization techniques and validating theoretical
models \citep{Changeat_2022, Mansfield_2021}.  The
temperature contrast between their permanent daysides and
nightsides, often exceeding 1000 K, provides a natural
laboratory for studying heat transport efficiency,
atmospheric circulation patterns, and the role of magnetic
fields in shaping planetary climates \citep{Bell_2018,
  2019ApJ...886...26T}.  Furthermore, the partially ionized nature of
their atmospheres allows for direct detection of elemental
species that remain locked in molecular form in cooler
planets, enabling precise abundance measurements that inform
planet formation theories and migration histories
\citep{Hoeijmakers_2019, Lothringer_2020}.

WASP-121b ($M_{\rm p}=1.16~M_J$, $R_{\rm p}=1.76~R_J$, 
$T_{\rm eq}\simeq 2358~\K$; \citealt{Evans_2016}; 
\citealt{2024AJ....168..231S}) orbits an F6V host star 
at $a=0.026~\au$ \citep{Evans_2016, Gaudi_2017}.
Its close-in orbit, combined with intense stellar 
irradiation, makes it an archetypal UHJ for studying 
atmospheric escape in a regime where stellar gravity and 
orbital dynamics significantly modulate the outflow 
morphology \citep{2025Natur.639..902S, 
2025NatCo..1610822A, 2024A&A...692A.230C}.


Recent observational advances have revealed remarkable
atmospheric details. The most extensively studied include
WASP-121b ($T_{\rm eq}\simeq 2358~\K$;
\citealt{Evans_2016}), WASP-76b
\citep[e.g.][]{Ehrenreich_2020}, WASP-189b
\citep[e.g.][]{Lendl_2020}, and the hottest known exoplanet
KELT-9b ($T_{\rm eq}>4000~\K$; \citealt{Gaudi_2017,
  Hoeijmakers_2019}).  High-resolution cross-correlation
spectroscopy (HRCCS) has identified species including atomic
metals such as \ion{Fe}{i}, \ion{Fe}{ii}, \ion{Na}{i},
\ion{V}{i} \citep[][]{Seidel_2019, Borsa2021}, along with
molecular signatures of CO and \chem{H_2O}
\citep{2022A&A...661L...6Y, 2023MNRAS.525.2985R}. These
observations frequently reveal temperature inversion layers
in dayside atmospheres \citep[][]{Sheppard_2017,
  Arcangeli_2019}, supersolar metallicities
\citep[][]{Changeat_2022, Pelletier_2025}, and significant
line broadening indicative of rapid rotation and atmospheric
dynamics, while nightside detections remain challenging due
to cooler temperature \citep{Stevenson_2017}.


General Circulation Models (GCMs) have served as the
cornerstone of UHJ atmospheric theory for over a decade,
successfully reproducing large-scale features such as
equatorial jets and day-night temperature contrasts
\citep[][]{Showman_2009, Lee_2022}.  However, these models
largely rely on the vertical hydrostatic approximation and
Local Thermodynamic Equilibrium (LTE), assumptions that
could break down in the upper atmospheric layers where
high-resolution spectroscopic observations are most
sensitive. 
The hydrostatic equilibrium assumption becomes invalid above
pressures of $\sim 10^{-5}~{\rm bar}$, precisely the regions
where atomic metal lines form and where atmospheric escape
processes dominate. 
Furthermore, GCMs typically employ simplified chemical
networks or equilibrium chemistry prescriptions that cannot
fully capture the non-equilibrium photochemistry and
ionization dynamics prevalent in the
thermosphere. 
This limitation is compounded by numerical dissipation and
coarse vertical resolution (often $\sim 40$ layers spanning
the entire atmosphere), which artificially suppress
vertically propagating modes and underestimate temperature
gradients in the uppermost
layers. 
Consequently, while GCMs excel at modeling the deep
atmospheres, they are not fully adequate to interpret the
spectral signatures that originate from the dynamic,
non-hydrostatic thermosphere where the impact by the host
star, including heating, photochemistry, and hydrodynamic
escape, collectively determine the observable atmospheric
signature.


To address these limitations, we present a series of
consistent three-dimensional UHJ simulations that couple
non-equilibrium thermochemistry, ray-tracing radiation, and
hydrodynamics using the Kratos GPU-accelerated simulation
system \citep{2025ApJS..277...63W, 2025arXiv250404941W}.
This approach solves the hydrodynamic equation using
higher-order Godunov method, conserving the extensive
quantities (mass, momentum, energy, and chemical species)
across a three dimensional (3D) mesh.  The Kratos framework
leverages mixed-precision arithmetic on heterogeneous GPU
architectures to achieve computational speeds $50-100\times$
faster than comparable CPU-based codes, enabling real-time
evolution of chemical networks comprising over 30 species
and $\sim 190$ reactions under full non-LTE conditions with
GPU-optimized semi-implicit method with adaptive step size
and error control \citep{2025arXiv250404941W}. We implement
a comprehensive photochemical model that includes XUV-driven
heating, ionization balance, and radiative cooling from
metal species, while the hydrodynamic solver captures
transonic outflow characteristics of atmospheric escape.
This multi-physics treatment allows us to simultaneously
model the transition from the dense lower atmosphere (where
GCMs are good at modeling) to the rarefied upper atmosphere
where transmission observations could probe, providing a
unified framework to interpret high-resolution transmission
spectroscopy while predicting the outflow hydrodynamics and
thermochemistry behaviors relatively accurately.

This paper is structured as follows. In \S\ref{sec:methods},
we provide a detailed description of the numerical
framework, including the setup of our 3D hydrodynamic
simulations, the comprehensive non-equilibrium
thermochemical network, and the methodology for generating
synthetic transmission spectra. \S\ref{sec:results} presents
the analyses of the fiducial simulation, characterizing the
structure, dynamics, and thermochemical composition of the
modeled outflowing atmosphere, and linking these features
directly to observable spectral
signatures. \S\ref{sec:results-para} systematically explores
the influence of key physical parameters, especially stellar
high-energy flux and stellar wind characteristics, on the
atmospheric escape process and its observational
diagnostics. Finally, \S\ref{sec:discussion} concludes our
major findings, discusses their implications for the
interpretation of current and future observations of UHJs.

\section{Methods}
\label{sec:methods}

The numerical framework employed in this work to model the
ultra-hot Jupiter WASP-121b is implemented using the
GPU-accelerated \kratos{} simulation system
\citep{2025ApJS..277...63W, 2025arXiv250404941W}. This
system integrates modules for consistent non-equilibrium
thermochemistry and ray-tracing radiative transfer, with
algorithms specifically optimized for Graphics Processing
Units (GPUs) to achieve the high computational performance
required for 3D global simulations. The underlying physical
mechanisms, including hydrodynamic escape driven by stellar
irradiation (calibrated to the stellar type) and the
interaction with stellar winds, are consistent with the
modeling approach previously developed for other evaporating
exoplanetary systems such as WASP-69b and WASP-107b
\citep{2021ApJ...914...98W, 2021ApJ...914...99W}. In the
following subsections, we provide a comprehensive summary of
the key methodological aspects, with the primary parameters
for our fiducial model detailed in
Table~\ref{table:fiducial-pars}.

\begin{deluxetable}{lr}
  \tablecolumns{2} 
  \tabletypesize{\scriptsize}
  \tablewidth{0pt}
  \tablecaption
  { Properties of the fiducial model for WASP-121b
    \label{table:fiducial-pars} }   
  \tablehead{
    \colhead{Item} &
    \colhead{Value}
  }
  \startdata
  Planet properties & \\
    $M_{\rm p}$ & $368~M_\oplus(\simeq 1.16~M_J)^*$ \\
    $R_{\rm p}$ & $19.6~R_\oplus(\simeq 1.76~R_J)^*$ \\
    $T_{\rm eq} $ & $2400~\K$ \\
  \\
  Stellar and orbit properties & \\
  $[ u_1, u_2 ]^{**}$ & $[0.33, 0.21]$ \\
  Orbital semi-major axis $a$ ${}^{***}$ & $0.026~\au$ \\
  Impact parameter $b$  ${}^{***}$ & $0.1~R_*$ \\
  \\  
  Simulation domain & \\
  Radial range & $19.6 \le (r/R_\oplus) \le \ 160$\\
  & $[1 \leq (r/R_{\rm p}) < 8.16\ ]$ \\
  Latitudinal range & $0\le\theta\le\pi/2$ \\
  Azimuthal range & $0\le\phi\le 2\pi$ \\
  Resolution $(N_{\log r}\times N_{\theta} \times
  N_{\phi})$ & 
  $128\times 32\times 128$ \\ 
  \\
  Radiation flux${}^\dagger$ [photon~$\cm~^{-2}~\s^{-1}$] & \\
  $2~\eV$ (IR/optical)  & $2.3\times 10^{21}$ \\3
  $4.9~\eV$ (Soft FUV)   & $2\times 10^{19}$ \\
  $6~\eV$ (FUV)   & $2\times 10^{18}$ \\  
  $12~\eV$ (LW)   & $1\times 10^{16}$ \\
  $20~\eV$ (Soft EUV ionizing H) & $4\times 10^{15}$ \\
  $60~\eV$ (Hard EUV)  & $1\times 10^{14}$ \\
  $0 .3~\keV$ (Soft X-ray) & $1\times 10^{14}$ \\    
  $3~\keV$ (X-ray) & $3\times 10^{13}$ \\   
  \\
  Initial abundances [$n_{\chem{X}}/n_{\chem{H}}$] & \\
  \chem{H_2} & 0.46\\
  He & 0.08\\
  \chem{H_2O} & $3.6 \times 10^{-4}$\\
  CO & $2.4 \times 10^{-4}$\\
  Na & $5 \times 10^{-6}$\\
  Mg & $6 \times 10^{-5}$\\
  Ca & $4 \times 10^{-6}$\\  
  Fe & $6 \times 10^{-5}$ \\
  Gr ${}^\ddagger$ & $1 \times 10^{-19}$ \\
  \\
  Dust/PAH properties${}^\ddagger$ & \\
  $r_\dust$ & $1~\ang$ \\
  $\sigma_\dust/\chem{H}$ (Effective specific cross section)
  & $3\times 10^{-35}~\cm^2$
  \enddata
  \tablecomments{
    $*$: Mass and radius of the Jupiter.\\
    $**$: Limb darkening parameters, see also
    \citet{2021AJ....161..294Y}.\\
    $***$: For planet orbital properties, see also 
      \citet{Evans_2016, Gaudi_2017, 2024AJ....168..231S}
    . \\
    $\dagger$: See \S\ref{sec:method-chem} for the
    representation of each energy bin. \\
    $\ddagger$: Very-small PAH and refractory graphitic
    grains could still survive sufficiently refractory to
    survive in the high-temperature regions ($T >
    1500~\K$; see also \S\ref{sec:method-chem}). 
  }
\end{deluxetable}

\subsection{Geometry and Boundary Conditions}
\label{sec:method-geo-kin}

We conduct our simulations on a spherical polar grid
$(r,\theta,\phi)$ that is centered on the planet and
co-rotates with its orbital motion. This coordinate choice
is essential to accurately capture the complex dynamics of
the atmospheric outflow, which is governed by the combined
influence of planetary gravity, stellar gravity, and orbital
forces (specifically the centrifugal and Coriolis forces
arising in the non-inertial rotating frame). The
computational domain spans the radial range from the
planetary surface at $R_{\rm p} = 19.6~R_\oplus$, out to an
outer boundary at $R_{\rm out} = 160~R_\oplus$. This
extensive radial coverage ensures that all critical physical
processes, from the dense lower atmosphere to the extended,
escaping exosphere, are contained within the simulation
volume. The initial mass density at the inner boundary is
set to $\rho_{\rm in} = 10^{-5}~\g~\cm^{-3}$, which
corresponds to $p_{\rm in} = 0.83~{\rm bar}$ with the
$T_{\rm eq}=2400~\K$ equilibrium temperature.  The
latitudinal ($\theta$) and azimuthal ($\phi$) domains cover
the upper hemisphere ($0 \le \theta \le \pi$,
$0 \le \phi \le \pi$), with the polar axis oriented
perpendicular to the orbital plane. The substellar point is
fixed at $(\theta, \phi) = (\pi/2, 0)$. To reduce
computational cost while preserving the dominant physical
symmetries, we model only the region above the orbital plane
and impose reflection symmetry across this plane for the
lower hemisphere.

The numerical grid is designed to resolve the large density
gradients expected in an escaping atmosphere. We employ a
logarithmically spaced radial grid to provide enhanced
resolution near the planetary surface, while the angular
grids in $\theta$ and $\phi$ are uniformly spaced.
Appropriate boundary conditions are applied at the domain
limits. At the inner boundary ($r = R_{\rm p}$), we impose a
reflecting condition, representing the impermeable planetary
surface (or the base of the simulated, quasi-isothermal
atmosphere). At the outer boundary ($r = R_{\rm out}$), an
outflow condition allows material to freely escape the
domain. The polar boundaries at $\theta = 0$ are treated as
polar wedges to avoid coordinate singularities.  To mitigate
the restrictive Courant-Friedrichs-Lewy (CFL) timestep
limitation imposed by the convergence of azimuthal grid
lines near the poles, we implement a mesh coarsening
technique in the $\phi$-direction at high latitudes
\citep[see e.g.][] {2019PASJ...71...98N,
  2019MNRAS.484.3307M}. This approach maintains the
conservation of mass, momentum, and energy while
significantly improving computational efficiency.

For our fiducial model of WASP-121b, we initialize the
atmosphere with an isothermal density profile corresponding
to an equilibrium temperature of $T_{\rm eq} = 2400~\K$,
consistent with its orbital distance of $a = 0.026~\au$ from
its F6V host star. Planetary gravity is treated as a point
mass $M_{\rm p} = 368~M_\oplus$ located at the origin of the
spherical polar coordinate system. The host star, positioned
outside the simulation domain at coordinates corresponding
to the orbital distance, influences the system through three
primary effects, (1) its gravitational pull, (2) the
inertial forces (centrifugal and Coriolis) in the
co-rotating frame, and (3) the injection of radiative flux
and stellar wind material. Given the short orbital period,
we further assume WASP-121b is tidally locked on a circular
orbit, thus the rotational angular frequency of the
simulation frame is identical to the orbital counterpart.

\subsection{Non-LTE Thermochemistry}
\label{sec:method-chem}

The intense high-energy radiation from the host star drives
the atmosphere of WASP-121b far from local thermodynamic
equilibrium (LTE). To model this, we discretize the stellar
spectral energy distribution (SED) into eight representative
energy bins, each responsible for distinct photochemical
processes:
\begin{enumerate}
\item $2~\eV$ (as a representative bin for $h\nu < 4.8~\eV$
  photons): Infrared, optical, and near-ultraviolet
  radiation, responsible for continuum heating.
\item $4.9~\eV$ (for $4.8~\eV < h\nu < 5.14~\eV$): Soft
  far-ultraviolet (FUV) photons that can destroy the neutral
  helium in the metastable triplet state \ext{He}
  ($I_{\ext{He}} = 4.8~\eV$).
\item $6~\eV$ (for $5.14~\eV < h\nu < 7.9~\eV$): Soft
  far-ultraviolet (FUV) photons capable of ionizing sodium
  ($I_{\rm Na} = 5.14~\eV$).
\item $12~\eV$ (for $7.9~\eV < h\nu < 13.6~\eV$, including
  the Lyman-Werner (LW) band $11.2~\eV < h\nu < 13.6~\eV$):
  \response{this bin covers the full $7.9$--$13.6~\eV$
  range, including the \lya line at $10.2~\eV$ and the}
  LW band photons, primarily driving the
  photodissociation of \chem{H_2} and CO, and can also
  ionize key tracing metals like iron (Fe), magnesium (Mg),
  and calcium (Ca).
\item $20~\eV$ (for $13.6~\eV < h\nu < 24.6~\eV$): Soft
  extreme-ultraviolet (EUV) photons that ionize atomic and
  molecular hydrogen (H and \chem{H_2}).
\item $60~\eV$ (for $24.6~\eV < h\nu < 0.1~\keV$): Hard EUV
  photons that additionally ionize helium (He).
\item $0.3~\keV$ (for $0.1~\keV < h\nu < 1~\keV)$: Soft
  X-ray photons.
\item $3~\keV$ (for $h\nu > 1~\keV$): Hard X-ray photons.
\end{enumerate}
The incident flux $F$ in photon number per unit area per
unit time. In $h\nu < 13.6~\eV$ energy bins, photon fluxes 
are adopted according to the F6V super-solar metallicity 
host star WASP-121 \citep{2024AJ....168..231S}. For 
$h\nu > 13.6~\eV$ energy bins, the EUV and X-ray fluxes are
calibrated to the data provided in \citet{2024A&A...692A.230C}. 
Fluxes of all relevant  bands are listed in 
Table~\ref{table:fiducial-pars}.

The simulations self-consistently couple ray-tracing
radiative transfer for these energy bins with hydrodynamics
and a comprehensive non-equilibrium thermochemical network
using the \kratos{} multiphysics framework
\citep{2025ApJS..277...63W, 2025arXiv250404941W}. The
chemical network, building upon established models for
irradiated atmospheres \citep[e.g.,][]{2021ApJ...914...98W},
has been expanded to include species and reactions critical
for ultra-hot Jupiter conditions. It comprises 147 thermal
reactions, including radiative and dielectronic
recombination, collisional excitation/de-excitation, and
associated heating and cooling processes (from the UMIST
Astrochemistry Database; \citealt{2013A&A...550A..36M}),
plus 29 radiation-driven reactions with cross
  sections and energy injection per reaction sourced
  primarily from \citet{1996ApJ...465..487V}. The shielding
  effects of relevant radiation reactions are taken into
  account via the data in \citet{2023AAP...675A..25H}. The
network tracks 33 equivalent species, including the specific
internal energy, plus 32 chemical species: \neg{e}, H,
\chem{H^+}, \neg{H}, \chem{H_2}, \chem{H_2^+}, \chem{H_3^+},
He, \pos{He}, \ext{He}, O, \chem{O_2}, \pos{O}, OH,
\pos{OH}, \pos{H2O}, H2O, \pos{H3O}, C, \pos{C},
\chem{CH^+}, CH, CO, Fe, \pos{Fe}, Na, \pos{Na}, Ca,
\pos{Ca}, Mg, \pos{Mg}, and Gr (representing a trace
population of refractory graphitic dust grains). Here,
\neg{H} is included as a prospectively significant source of
continuum opacity in hot atmospheres \citep[see e.g.,][]
{2018ApJ...855L..30A, 2022A&A...668L...1J}. Initial
abundances, assuming a bulk metallicity approximately twice
solar \citep{2024AJ....168..231S}, are provided in
Table~\ref{table:fiducial-pars}. The associated stiff system
of ordinary differential equations for the reaction network
is solved efficiently using a semi-implicit integration
scheme optimized for GPU architectures
\citep{2025arXiv250404941W}.

The chemical reaction set shares its lineage with the
thermochemical networks presented in the exoplanetary escape
models of \citet{2021ApJ...914...98W,
  2021ApJ...914...99W}. Extensions specific to UHJ physical
conditions and observational tracers mainly include
additional metal tracers (Na, Fe, and Mg, Ca) required for
transmission spectroscopy diagnostics. Comparisons of the
network composition against the established networks of
\citet{2013Icar..226.1678K}, \citet{2011ApJ...737...15M},
and \citet{2017ApJ...851..150H} are provided in
Appendix~\ref{sec:appendix-reactions}. Non-equilibrium
cooling processes are included alongside the reaction
network. Molecular cooling from CO, H$_2$, H$_2$O, and OH is
implemented following the interpolation tables in
\citet{1993ApJ...418..263N} and \citet{2010ApJ...722.1793O}.
\response{ Five metal atomic line cooling processes from
  \chem{C^+}, C, and O, based on the data compiled in
  chapter~3 of \citet{2011piim.book.....D} (and references
  therein), are included, using the escape-probability
  estimation of \citet{1981ApJ...250..478K} with the
  line-center optical depth estimated via the Sobolev
  approximation (see also
  Appendix~\ref{sec:appendix-reactions}). \lya
  cooling is not included in the fiducial model: as stated
  in chapter~3 of \citet{2011piim.book.....D} and confirmed
  by Appendix~A of \citet{2018ApJ...860..175W}, \lya
  cooling alone is generally not sufficient to establish a
  steady state in photoevaporating planetary atmospheres,
  and the metal fine-structure lines are the more important
  coolants at the $\sim 10^3$~K temperatures prevailing in
  the outflow. We have verified this with an additional test
  run including \lya cooling (six atomic cooling
  processes in total) and otherwise identical setups; all
  diagnostics and the mass-loss rate agree with the fiducial
  model to within a few percent
  (Appendix~\ref{sec:appendix-reactions}). }

\begin{figure*}
  \centering
  \vspace{-0.5cm}  
  \hspace{-0.8cm}\includegraphics[width=0.33\linewidth]
  {\figdir/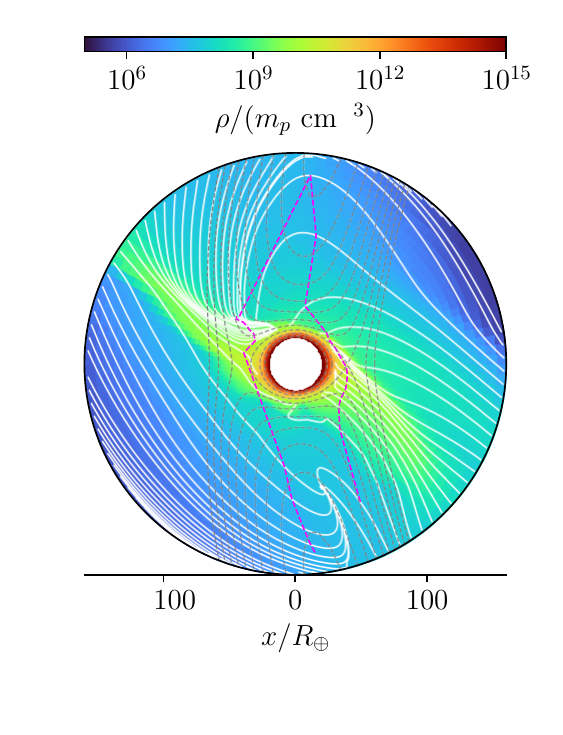}
  \hspace{-0.8cm}\includegraphics[width=0.33\linewidth]
  {\figdir/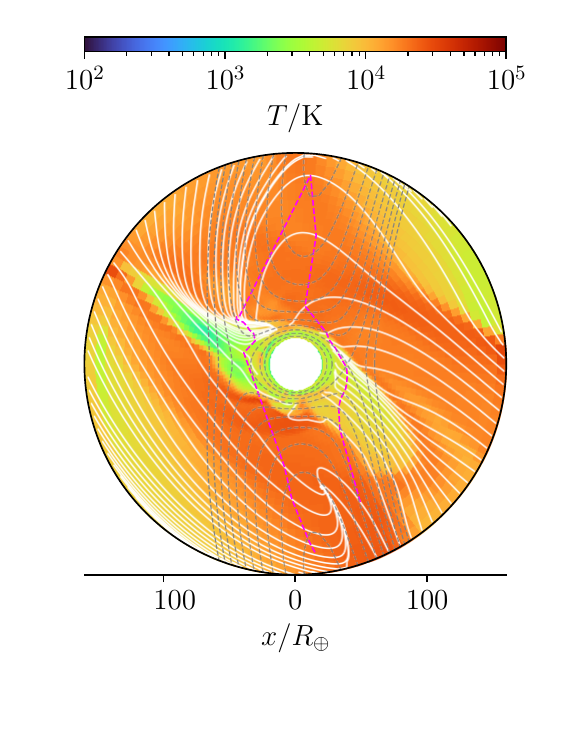}
  \hspace{-0.8cm}\includegraphics[width=0.33\linewidth]
  {\figdir/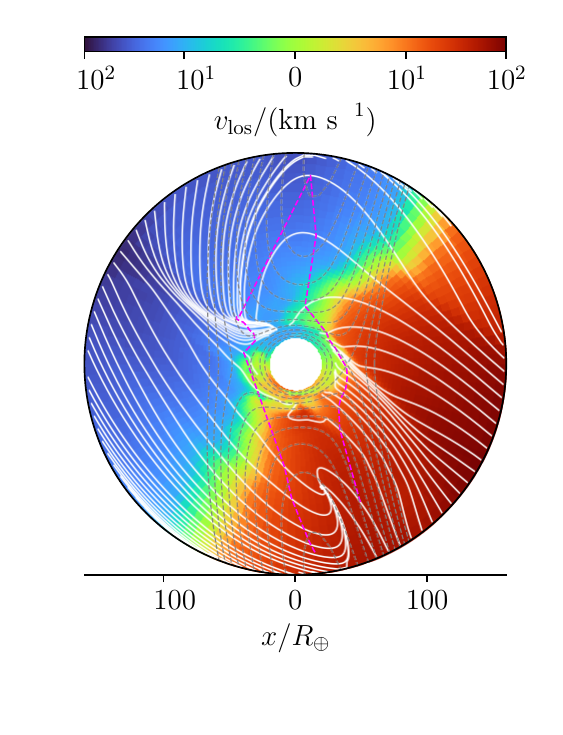}
  \\
  \vspace{-1.1cm}
  \hspace{-0.8cm}\includegraphics[width=0.33\linewidth]
  {\figdir/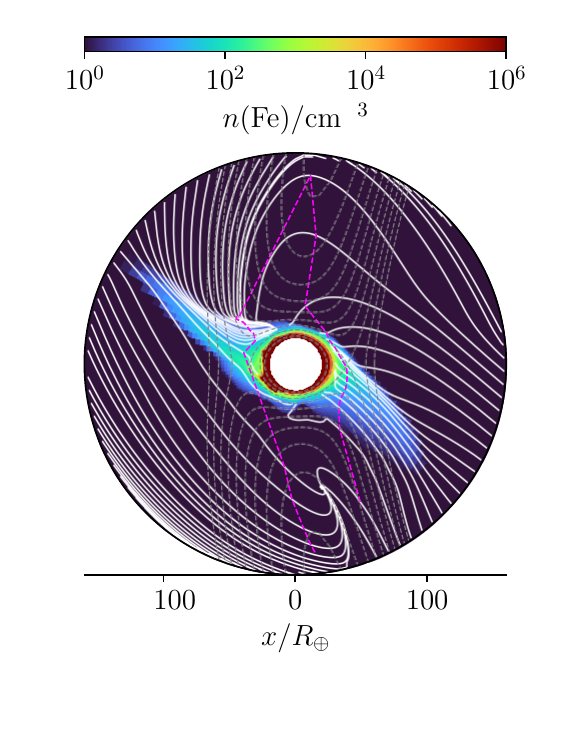}
  \hspace{-0.8cm}\includegraphics[width=0.33\linewidth]
  {\figdir/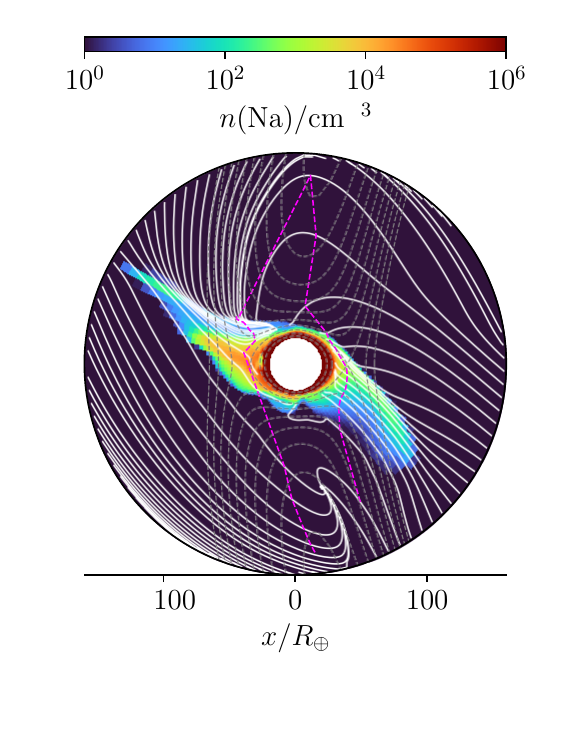}
  \hspace{-0.8cm}\includegraphics[width=0.33\linewidth]
  {\figdir/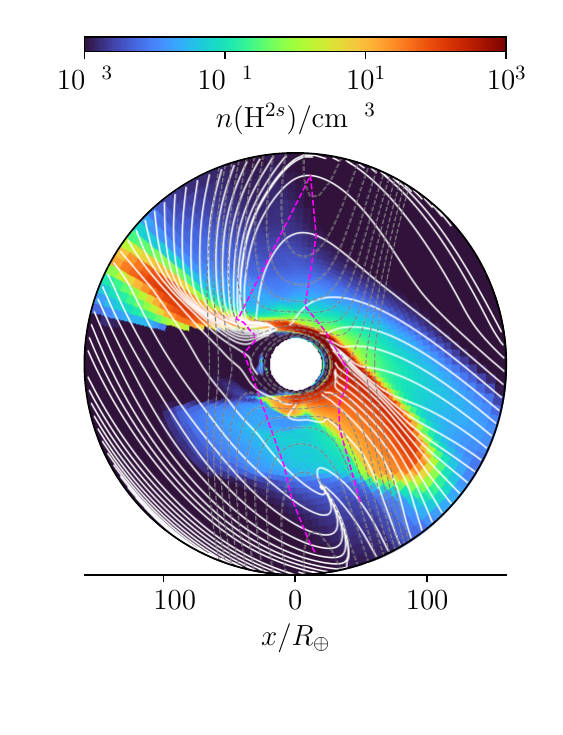}
  \\
  \vspace{-1.1cm}
  \hspace{-0.8cm}\includegraphics[width=0.33\linewidth]
  {\figdir/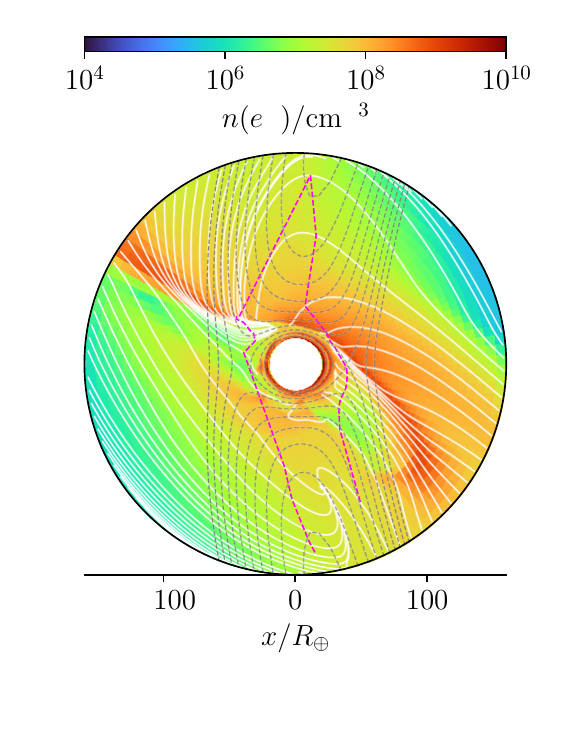}
  \hspace{-0.8cm}\includegraphics[width=0.33\linewidth]
  {\figdir/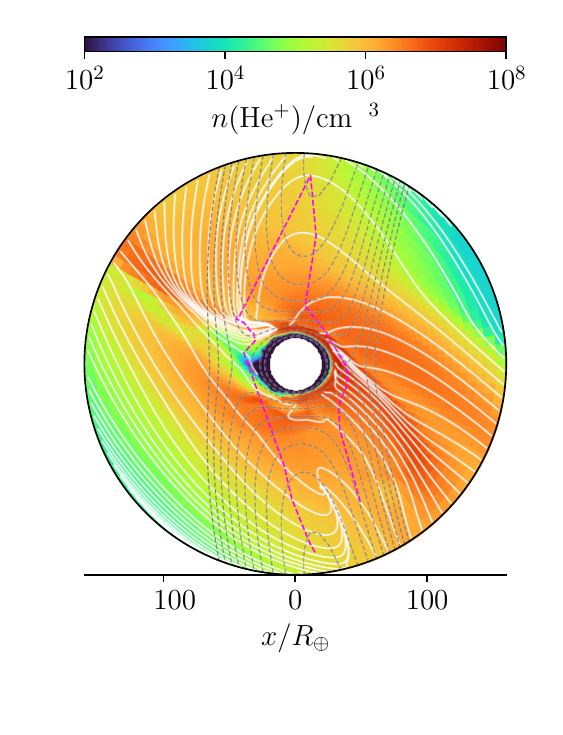}
  \hspace{-0.8cm}\includegraphics[width=0.33\linewidth]
  {\figdir/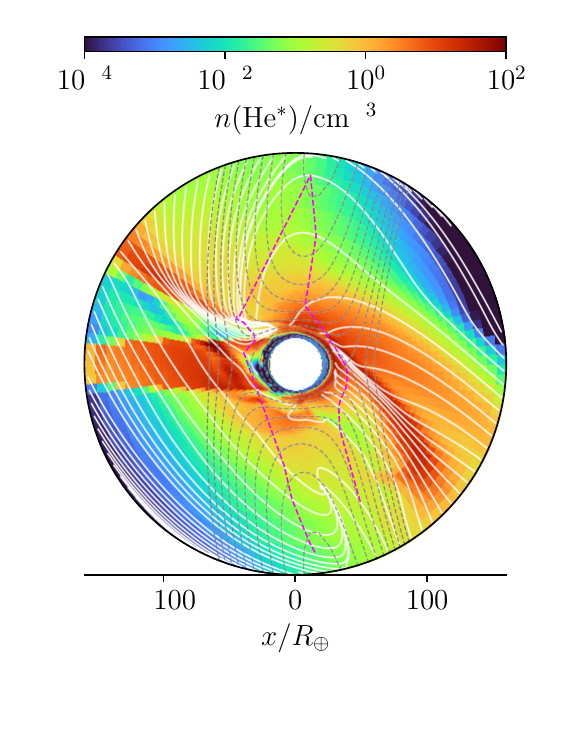}
  \vspace{-1.1cm}
  \caption{Equatorial slices of key hydrodynamic and
    thermochemical quantities from the fiducial simulation
    of WASP-121b. The panels show (from left to right, top
    to bottom): gas density $\rho$, temperature $T$,
    line-of-sight velocity in the laboratory frame
    $v_{\rm los}$, and the number densities of Fe, Na, 
    H$^{2s}$ (metastable neutral hydrogen on the $2s$ state, 
    the H$\alpha$ absorber), free electrons ($e^-$), He$^+$, 
    and \ext{He} (metastable neutral helium). The snapshot is the 
    temporal average over the final 10 hours over the $300$
    total simulated hours, after evolving the system through 
    a quasi-steady state (\S\ref{sec:method-chem}). 
    The spatial coordinates are in units
    of Earth radii ($R_\oplus$), and the central circle
    marks the nominal planetary radius ($R_{\rm p}$).
    Equipotential contours are shown in gray dotted
    lines, streamlines are indicated by white lines, and the
    sonic critical lines are overlaid as dashed magenta
    lines. The planet orbital motion is downwards, and the 
    host star is to the right. }
  \label{fig:slice-fid}
\end{figure*}

The inclusion of full non-LTE thermochemistry significantly
increases the computational cost. A typical 3D simulation
for WASP-121b requires $\sim 22$ hours of wall-clock time on
two NVIDIA RTX 4090 GPUs to simulate $\sim 200$ hours of
physical time. This duration is justified by the estimated
dynamical timescale,
\begin{equation}
  \label{eq:t-dyn-wasp121b}
  \tau_\dyn \sim \dfrac{G M_p}{c_s^3} \sim 30~{\rm hr} \times
  \left(\dfrac{M_p}{1.18~M_{\rm J}}\right)
  \left(\dfrac{T}{10^4~\K}\right)^{-3/2},
\end{equation}
which, despite potential shear instabilities, is sufficient
for the system to reach a quasi-steady state. We first run
the simulations for 100 simulated hours with hydrodynamics
only (turning off thermochemistry and radiation), letting
the hydrodynamic profiles to adjust to and settle down with
the effective equipotential contours, and then continue the
simulations for 200 simulated hours through the final
quasi-steady state.  To obtain representative conditions for
synthetic observations, we time-average all physical
quantities over the final 10 simulated hours of each run and
perform spectral synthesis on this averaged data.

\subsection{Synthetic Observations}
\label{sec:method-spec}

To directly compare our simulation results with
high-resolution spectroscopic observations, we generate
synthetic transmission spectra for selected absorption
lines. For a given chemical species $X$, the
wavelength-dependent optical depth $\tau$ along a specific
line of sight (LoS) at orbital phase $\Phi$ is calculated by
integrating the contributions from all velocity components
along the path:
\begin{equation}
  \label{eq:method-tau-los}
  \begin{split}
  \tau(\lambda; \Phi) & = \int_{\rm LoS} |\d \mathbf{x}|\,
  n(X;\mathbf{x}; \Phi) \\
  & \quad \times \sum_i \sigma_i[\lambda;
  \hat{n}_{\rm los}\cdot \mathbf{v}(\mathbf{x}; \Phi),
  T(\mathbf{x}; \Phi)]\ ,
  \end{split} 
\end{equation}
where $n(X; \mathbf{x}; \Phi)$ is the number density of the
species, and the sum runs over all relevant spectral line
components (e.g., the three fine-structure lines of the
\ext{He} $10830~\ang$ triplet). The cross-section $\sigma_i$
for each component is modeled as a Voigt profile, accounting
for natural broadening, thermal Doppler broadening at the
local temperature $T(\mathbf{x})$, and Doppler shifts
induced by the projected bulk velocity
$\hat{n}_{\rm los}\cdot \mathbf{v}(\mathbf{x})$ along the
LoS.

The relative extinction (i.e., absorption depth) at
wavelength $\lambda$ is then obtained by integrating the
attenuated stellar flux over the visible stellar disk:
\begin{equation}
  \label{eq:method-extinction}
  \epsilon(\lambda;\Phi) = 1 - \int \d\Sigma\, S(\lambda)
  \exp[-\tau(\lambda;\Phi)]\ .
\end{equation}
Here, $S(\lambda)$ is the normalized stellar surface
brightness profile, which incorporates limb-darkening and
rotational broadening effects appropriate for the host
star. This calculation is repeated for multiple LoS
intersecting the planet's atmospheric annulus at each time
step during the transit. The resulting collection of
$\epsilon(\lambda; \Phi)$ across all phases and LoS
constitutes our synthetic transmission spectrum, which can
be compared directly to observed data to validate the
physical model and interpret kinematic and chemical
features.   While the transmission spectra of
  neutral Na and Fe are relatively straightforward to
  implement with chemistry calculations, the calculations
  for H$\alpha$ and He $10830~\ang$ deserve some further
  discussions. 

\subsubsection{H$\alpha$ Absorber Population}

The lower level of the H$\alpha$ transition
($\chem{H}^{2s}$) is populated in the thermosphere through a
combination of radiative and collisional processes.  In the
fiducial simulation described in \S\ref{sec:method-chem},
the $\chem{H}^{2s}$ population is determined by
recombination of H$^+$ and charge exchange between H and
H$^+$, which are significantly faster than direct
collisional pumping.

To incorporate the effect of \lya radiative pumping
into the $\chem{H}^{2s}$ population, we perform
post-processing calculations.
One notices that, when the line-center optical depth reaches
$\tau\sim 10^5-10^6$, an excessive number
($\sim 10^{10}-10^{12}$) of scattering events per photon is
demanded for acceptable signal-to-noise ratios. As the full
\lya Monte-Carlo calculations is not viable given the
typical optical depths, the mean intensity of \lya in
the thermospheric region is estimated from the optical depth
along the line of sight to the host star with the
Monte-Carlo radiative transfer results calibrated by
\citet{2017ApJ...851..150H} for the high-optical-depth
regime. The $\chem{H}^{2s}$ population fraction is then
computed by balancing \lya pumping, collisional
$2s \leftrightarrow 2p$ level crossing (using the rate
coefficients of \citep{2006agna.book.....O}), and
recombination/charge-exchange processes.  \response{This
calibrated estimate is an approximation to a full
\lya radiative transfer treatment, which is not
currently viable given the extreme line-center optical
depths of the thermosphere. We note that full Monte-Carlo
\lya radiative transfer calculations in the
literature \citep{2019ApJ...884L..43G,2021A&A...645A..22Y,
2022A&A...661L...6Y,2017ApJ...851..150H} generally find
\lya pumping to be the dominant contributor to the
H($2p$)/H($2s$) population; our calibrated approach
preserves this dominance through the high-optical-depth
calibration, but we caution that it does not replace a
self-consistent radiative transfer solution.} For comparison,
we also compute the H$^{2s}$ number density and the H$\alpha$
spectra for the fiducial model using an alternative
$\chem{H}^{2s}$ population scheme that omits \lya
pumping (see Appendix~\ref{sec:appendix-ha-old}).  This
alternative method yields very similar line profiles and
comparable amplitudes to the \lya-informed approach,
as discussed in \S\ref{sec:H_alpha}.


\subsubsection{Metastable Helium Population}

The metastable helium He ($2^3S$) population (denoted by
\ext{He}), which produces the He~10830~\ang\ absorption, is
computed within the live thermochemical network. The
formation channels include recombination of He$^+$ (the
predominant channel; see also \citealt{2018ApJ...855L..11O,
  2021ApJ...914...98W, 2021ApJ...914...99W}) and collisional
excitation by free electrons. Destruction pathways comprise
spontaneous radiative decay (timescale
$\simeq 2.2~{\rm hr}$), collisional de-excitation with free
electrons and neutral hydrogen atoms, and direct
photoionization by photons with energy $h\nu > 4.8~\eV$ (the
ionization threshold from the metastable triplet state),
with cross sections adopted from \citet{2018ApJ...855L..11O}
and \citet{2021ApJ...914...98W}.  We have also conducted a
test simulation that explicitly includes He$^{2+}$ and the
associated He$^+$ photoionization and He$^{2+}$
recombination reactions (Appendix~\ref{sec:appendix-hepp}),
confirming that, while He$^{2+}$ dominates over He$^+$ in
the diffuse outer regions of the simulation domain, those
regions do not contribute significantly to the
He~10830~\ang\ optical depth ($<10\%$ change in the
equivalent width).

\section{Results: Outflowing Atmospheres and Observables}
\label{sec:results}

After evolving the fiducial model for $300$ simulated hours
through the quasi-steady state, this section presents the
principal findings from our fiducial three-dimensional
simulation, which models the interactions between the
extended atmosphere of WASP-121b and its host star through
radiative transfer and stellar wind forcing. In general, our
results reveal that the outer atmospheric structure is
dominated by a global, supersonic outflow that is sculpted
by orbital motion and the Coriolis force into two spiral
arms, as illustrated in Figures~\ref{fig:slice-fid} and
\ref{fig:render-fid}, and explained schematically in
Figure~\ref{fig:scheme}.  The synthetic transmission spectra
generated from these simulations demonstrate that specific
chemical tracers, each surviving in different atmospheric
reservoirs, serve as probes of distinct atmospheric layers
and physical processes, from deep circulation to
high-altitude dynamics of atmospheric escape.

\subsection{Morphology and Dynamics of the Planetary
  Atmosphere}
\label{sec:morphology}

\begin{figure*}
  \centering
  \includegraphics[trim={6.5cm 0 0 1.5cm},clip,
  width=0.49\linewidth]
  {\figdir/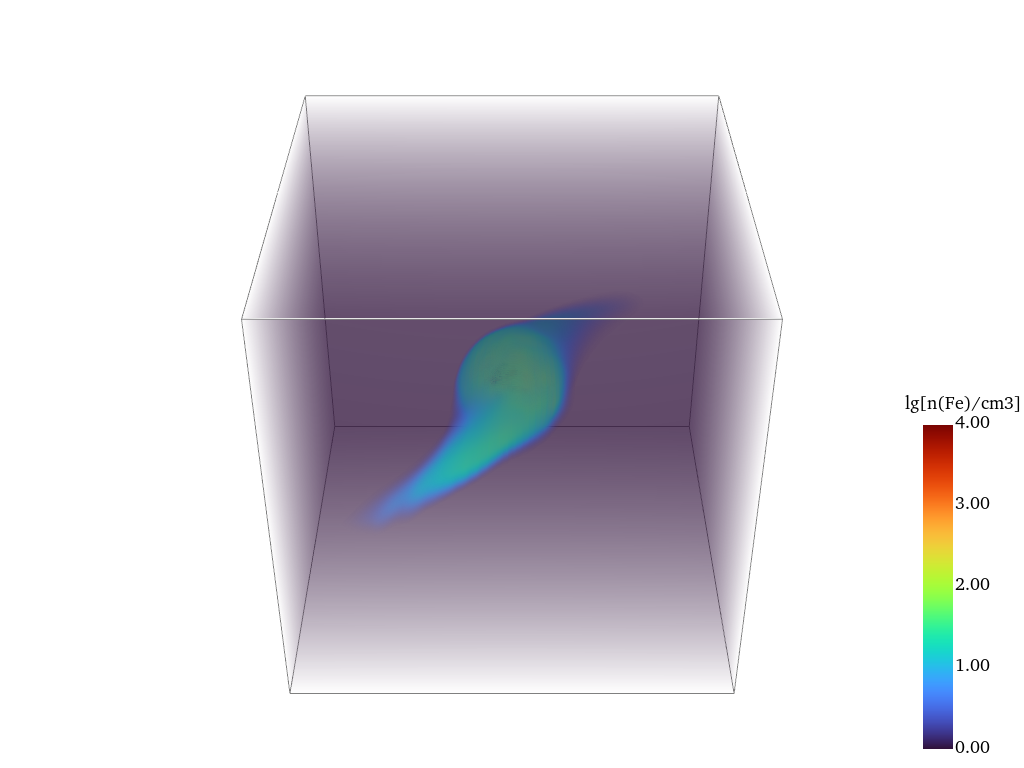}
  \includegraphics[trim={6.5cm 0 0 1.5cm},clip,
  width=0.49\linewidth] 
  {\figdir/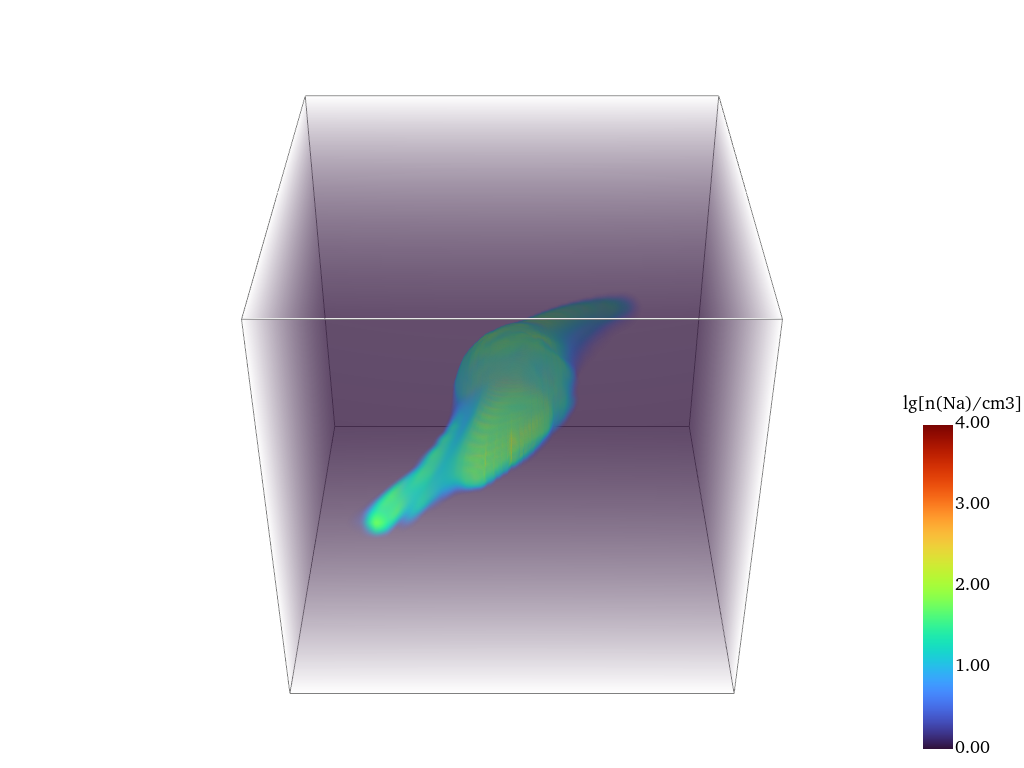} 
  \\
  \includegraphics[trim={6.5cm 0 0 1.5cm},clip,
  width=0.49\linewidth]
  {\figdir/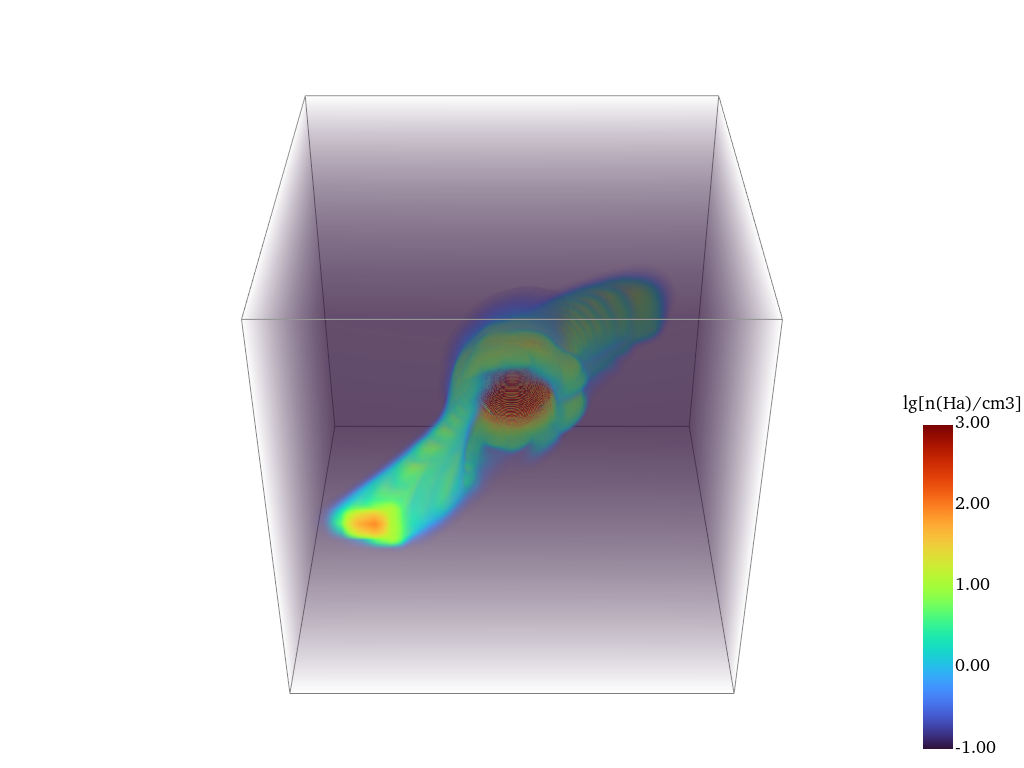}
  \includegraphics[trim={6.5cm 0 0 1.5cm},clip,
  width=0.49\linewidth] 
  {\figdir/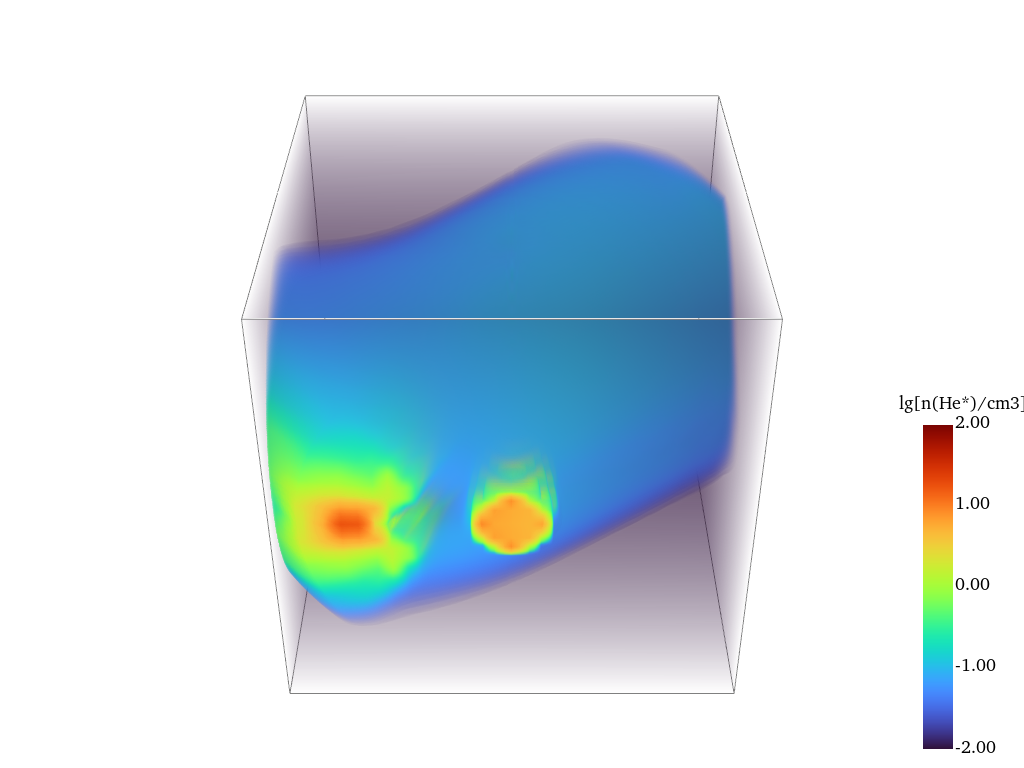}   
  \caption{Colormapped volume rendering for the 10-based
    logarithms of number densities (in $\cm^{-3}$) for four
    key tracing species (top left: Fe; top right: Na; bottom
    left: H$\alpha$; bottom right: \ext{He}). Note that the 
    dynamical ranges of colormaps are different. The host
    star is on the far side from the reader, and the planet
    moves from left to right. The rendering boxes have the
    same sizes ($240~R_\oplus$ along every dimension).  The
    high-abundance region of \ext{He} in the planet shadow
    is clearly seen as a cylinder. }
  \label{fig:render-fid}
\end{figure*}

From both Figure~\ref{fig:slice-fid} for the equitorial
slices and Figure~\ref{fig:render-fid} showing the volume
rendering, one can clearly identify that the atmospheric
dynamics of the model planet is strongly influenced by the
host star gravitational field and the orbital motion, which
together introduce significant centrifugal and Coriolis
forces in the co-rotating reference frame. The planetary
interior and lower atmosphere, as traced by the density and
chemical species maps, conform closely to the calculated
equipotential surfaces, adopting a predominantly ellipsoidal
shape elongated toward the host star. This elongation is a
direct consequence of the substantial rotational and tidal
deformation experienced by the planet in its extremely
close-in orbit ($a=0.026~\au$).

\begin{figure}
  \centering
  \includegraphics[width=0.7\linewidth]
  {\figdir/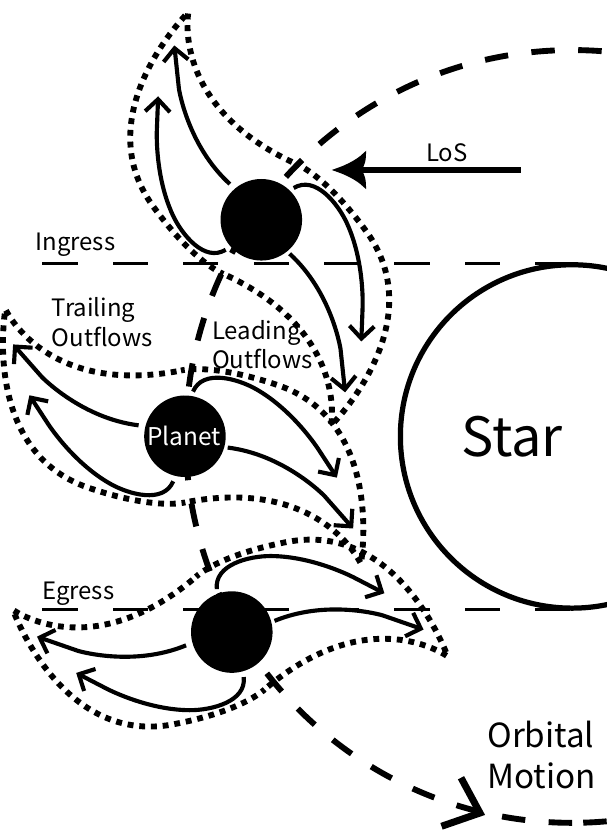} \\
  \caption{Schematic illustration of the transit geometry
    and the origin of asymmetric velocity shifts. The planet
    (black solid circle) moves from top to bottom. The
    leading (redhifted) and trailing (blueshifted) spiral
    arms are shaped by both the spilling through the
    Lagrangian points and the Coriolis force
    (\S\ref{sec:morphology}). The LoS during ingress and
    egress samples different projections of the arm
    velocities, leading to the observed asymmetry in the
    transmission spectrum.}
  \label{fig:scheme}
\end{figure}

\subsubsection{Lower altitudes and atmospheric advection}
\label{sec:morph-advection}

The equipotential contours overlaid on the hydrodynamic
profiles reveal that the Roche lobe, the region where
planetary gravity dominates over stellar tidal forces,
extends only a few $R_\oplus$ beyond the planetary body
itself. The characteristic size of this region is given by
the Hill radius,
\begin{equation}
  \label{eq:r-hill}
  \begin{split}
    R_{\rm Hill}
    & \simeq a \left(\dfrac{M_{\rm p}}{3M_*}\right)^{1/3}
      \simeq 38 R_\oplus\times
      \left( \dfrac{a}{0.026~\au} \right) \\
    & \times
      \left(\dfrac{M_{\rm p}}{1.18 M_{\rm J}}\right)^{1/3}
      \left( \dfrac{M_*}{1.35 M_\odot} \right)^{-1/3}\ .
  \end{split} 
\end{equation}
When compared to the measured planetary radius of
$R_{\rm p} \simeq 21~R_\oplus$, the ratio
$R_{\rm Hill}/R_{\rm p} \lesssim 2$ indicates that the
domain of effective planetary gravitational control extends
to less than twice the planetary radius from its
center. The escape velocity from the planet surface could
be estimated as,
\begin{equation}
  \label{eq:v-esc}
  v_{\rm esc} \simeq \left[ 2 G M_{\rm p}\left(R_{\rm p}^{-1}
      - R_{\rm Hill}^{-1} \right) \right]^{-1/2}
  \simeq 30~\km~\s^{-1}\ .
\end{equation}
Such a relatively low threshold considerably eases the
escape of high-velocity streams. Consequently, gaseous
material that escapes beyond approximately $1.8~R_{\rm p}$
undergoes a transition from being primarily bound by the
planetary gravity to being governed by orbital kinematics
and the stellar gravitational potential.

\begin{figure*}
  \centering
  \includegraphics[width=0.48\linewidth]
  {\figdir/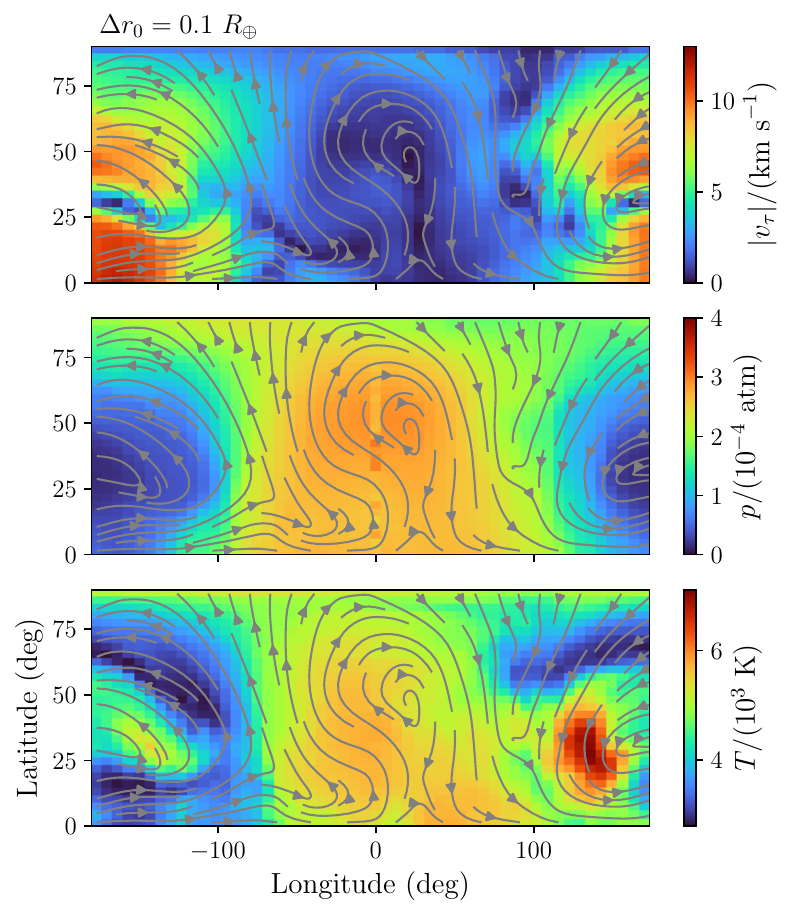}
  \quad
  \includegraphics[width=0.48\linewidth]
  {\figdir/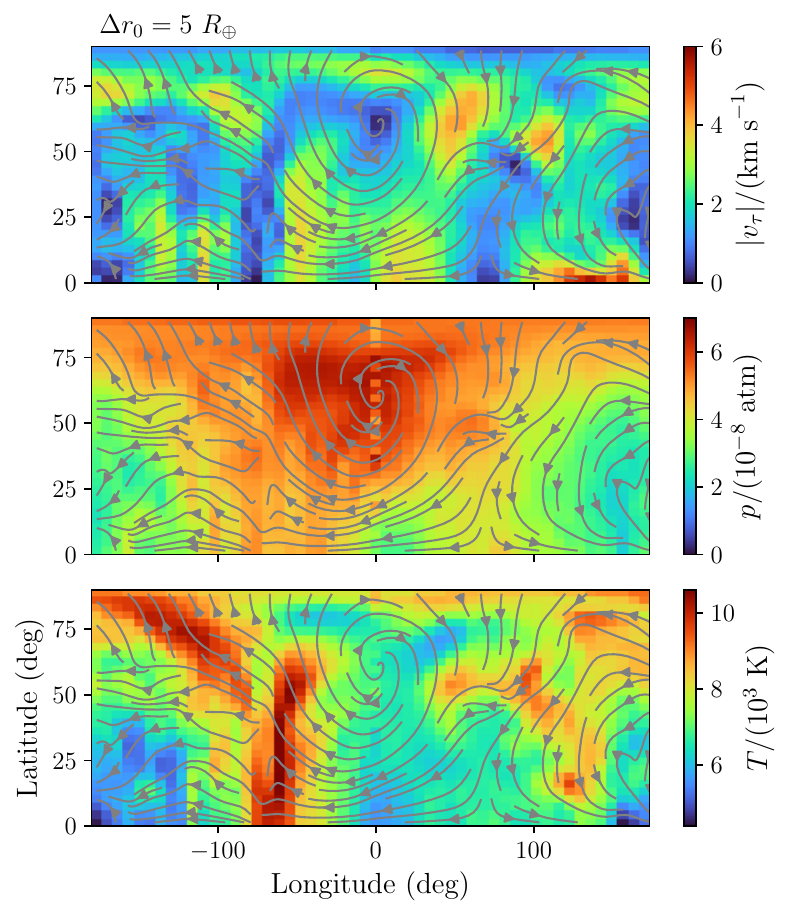} 
  \caption{Instantaneous velocity fields in streamlines and
    arrows, and hydrodynamic quantities in colormaps
    (tangential wind speed in the top row, pressure in the
    middle row, and temperature in the bottowm row), showing
    two representative equipotential surfaces from the
    fiducial simulation. Left column presents the surface
    near the planetary surface ($\Delta r_0 = 0.1~R_\oplus$
    above the substellar radius), illustrating the
    substellar anticyclone and the antistellar cyclone,
    along with prograde super-rotation near the
    equator. Right column presents the equipotential surface
    at $\Delta r_0 = 5~R_\oplus$, where the retrograde winds
    are predominantly deflected by the Coriolis force on the
    outflow.}
  \label{fig:surface-fid}
\end{figure*}

The hydrodynamic velocity fields presented in
Figure~\ref{fig:surface-fid} illustrate these contrasting
flow behaviors across two representative equipotential
surfaces.  It is noted that the reflecting inner radial
boundary condition applied at the planetary surface may not
fully capture detailed rotational dynamics or convective
motions originating from the deep planetary
interior. Therefore, results pertaining to the innermost
atmospheric layers within a few hundred kilometers of the
inner boundary should be interpreted primarily in
qualitative terms.  We also note that the flow patterns are
time-dependent and may not converge to a strict steady state
over long timescales; the physical quantities shown
represent instantaneous snapshots from a dynamically
evolving system.

At the lower equipotential surface, located only
$\Delta r_0 = 0.1~R_\oplus$ above the planetary radius at
the substellar point, the flow structure is characterized by
a large-scale clockwise anticyclone centered near the
substellar longitude, and a counter-clockwise cyclone
situated around the antistellar longitude. It is noticed
that, although our simulation only covers the northern
hemisphere and assumes reflection symmetry across the
equatorial plane, the existence and locations of such
cyclone and anticyclone systems are not qualitatively
affected by this simplification, as they cannot exist across
the equator (because the directions of rotation are
different in the northern versus southern
hemispheres). These vortical systems are embedded within
broader prograde (west-to-east) zonal winds at the equator,
particularly pronounced near the antistellar point where the
cyclonic flow itself is prograde. These complex flow
patterns arise primarily from day-to-night pressure
gradients, which drive gas from the heated dayside toward
the cooler nightside. The resulting flow is then deflected
by the Coriolis force associated with the planet's tidally
locked spin, generating the observed anticyclonic and
cyclonic circulations. 

At a higher equipotential surface
located $\Delta r_0 = 5~R_\oplus$ above the substellar
surface, which is still within the planetary Roche lobe but
influenced by the expanding outflow, the flow morphology
undergoes a notable transition. While residual anticyclonic
and cyclonic structures remain visible, the dominant
transverse gas motion becomes systematically retrograde
(east-to-west).

The significant vertical shear in zonal wind direction, from
prograde flow at depth to retrograde motion at higher
altitudes, creates conditions conducive to the development
of Kelvin-Helmholtz instabilities (KHI).  In the fiducial
model, this shear layer is most pronounced in the night
hemisphere and near the day-night terminators, while still
exist in the day hemisphere (see the tangential velocity
panels compared in Figure~\ref{fig:surface-fid}). The
resulting turbulence can modulate the atmospheric structure
on timescales of hours to days, intermittently enhancing or
suppressing the local mass flux escaping the planetary
potential.  Such variations could, in principle, imprint
onto time-resolved transmission spectra, and current
observations qualitatively imply such variabilities
\citep[e.g.][]{2024ApJS..270...34C}.  Future observations
with higher cadences signal-to-noise ratios (SNRs), and
spectral resolutions, are required to quantitatively capture
these sub-transit variations in detail.

\begin{figure*}
  \centering
  \includegraphics[width=0.88\linewidth]
  {\figdir/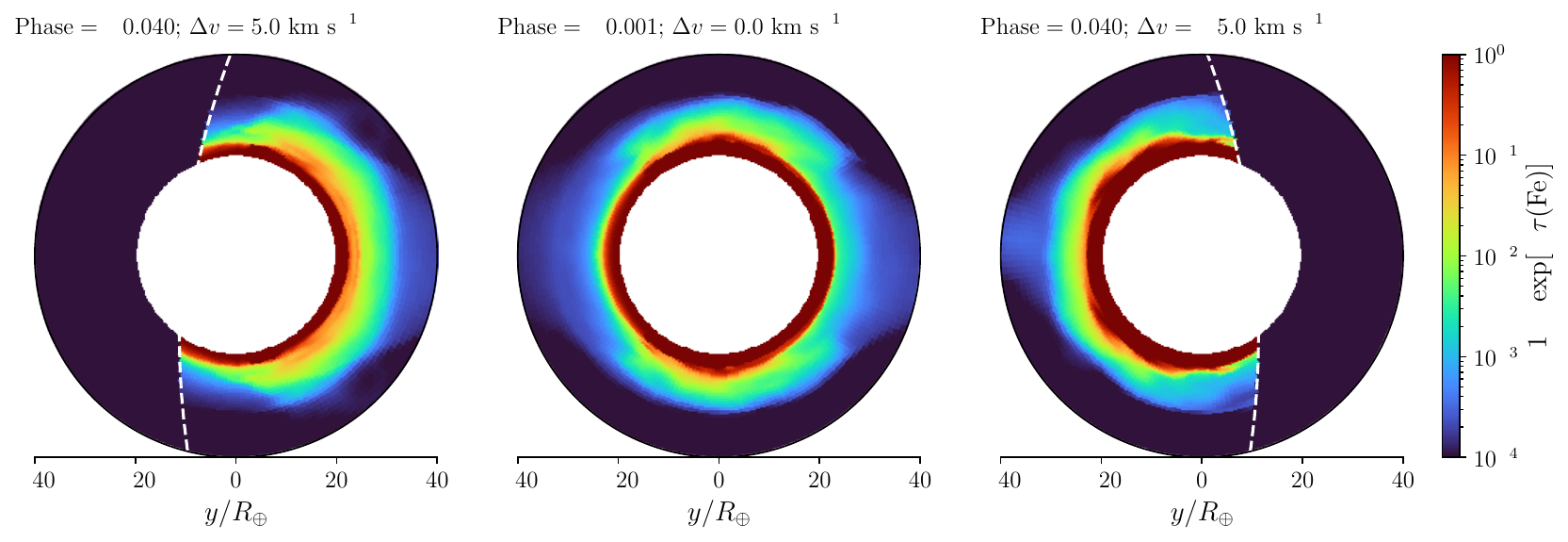} \\  
  \includegraphics[width=0.88\linewidth]
  {\figdir/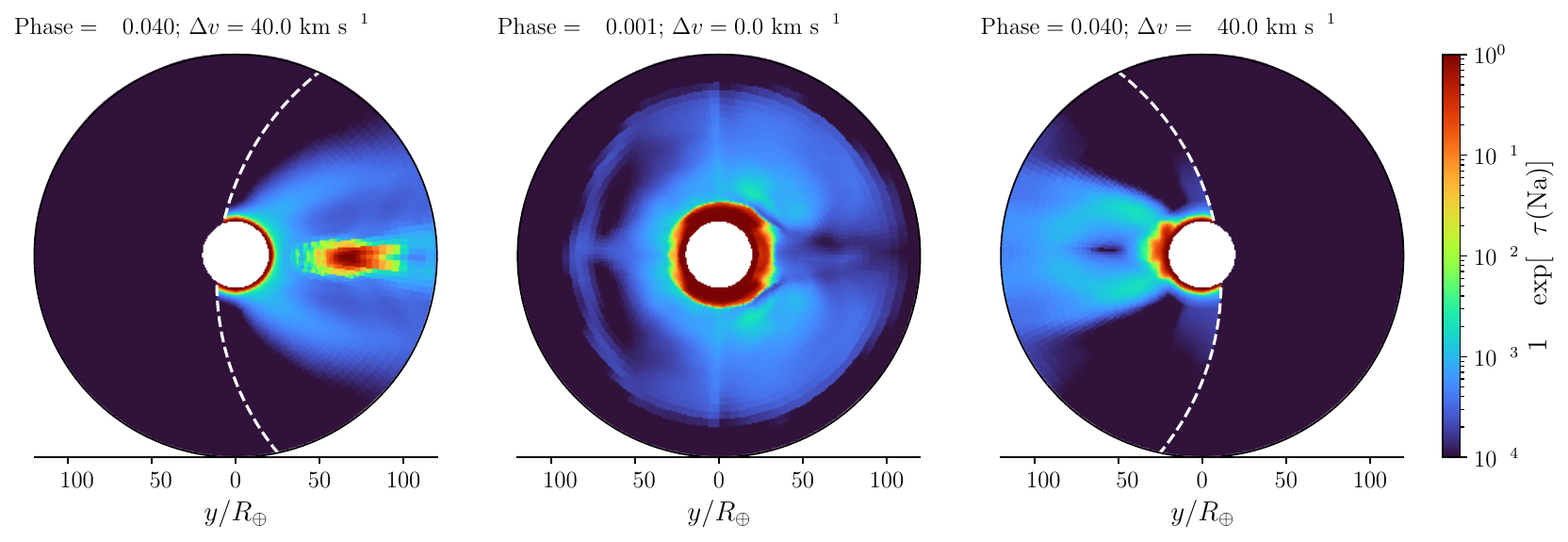} \\
  \includegraphics[width=0.88\linewidth]
  {\figdir/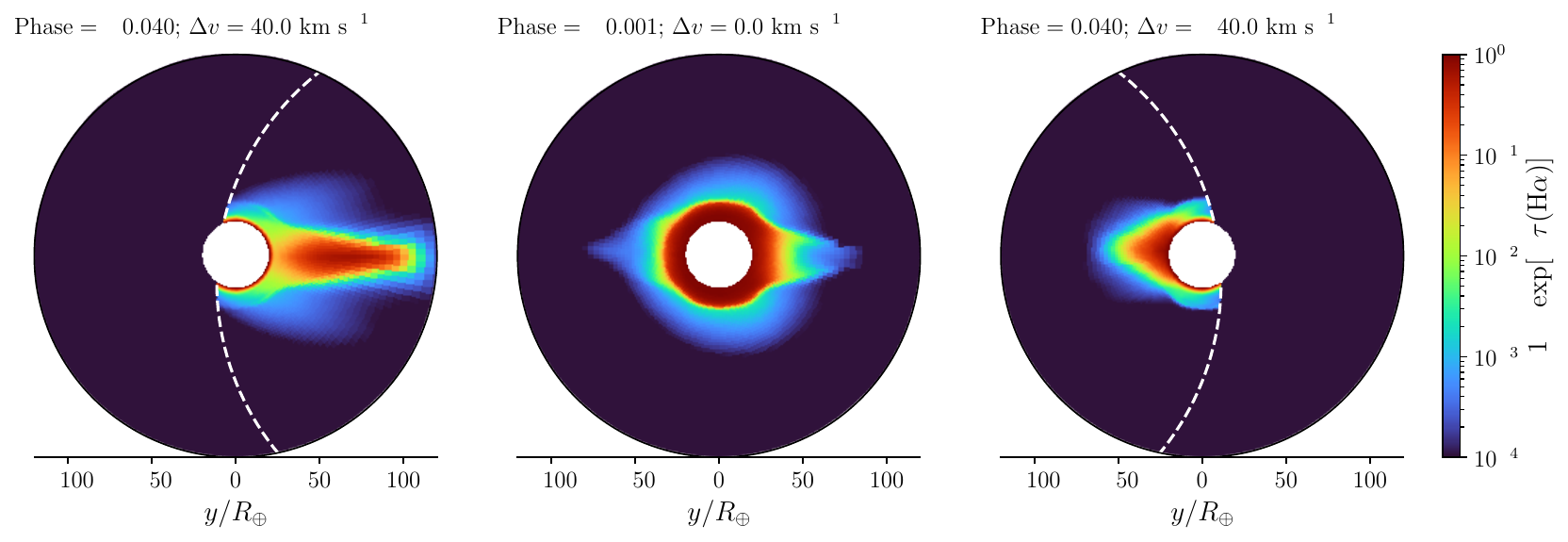} \\
  \includegraphics[width=0.88\linewidth]
  {\figdir/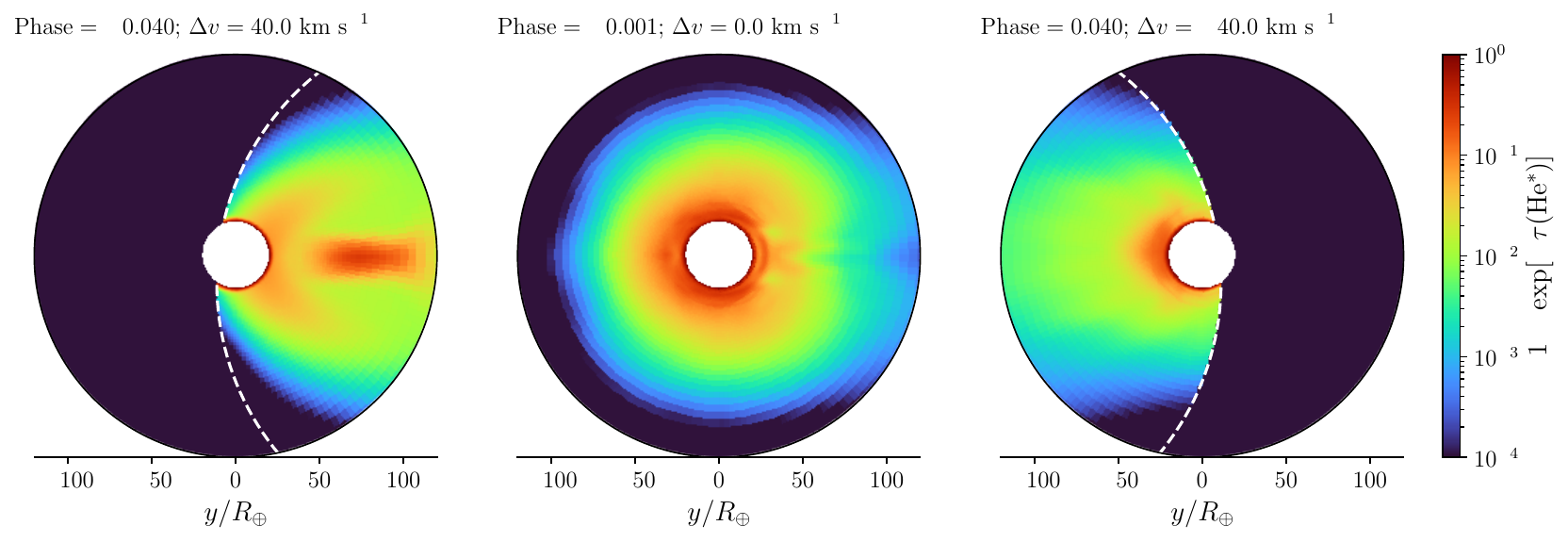} 
  \caption{Extinction intensities (quantified by
    $1-\e^{-\tau}$) at three different velocities and orbit
    phases (denoted at the top of each panel) for four key
    tracers from the fiducial simulation (denoted at the
    colobar in each row). White dashed circles indicate the
    projection of the host star, in which the extinction by
    planetary atmospheres are calculated.  The asymmetric
    velocity shifts and phase-dependent absorption depths
    trace the geometry and kinematics of the spiral arms and
    inner atmospheric layers.}
  \label{fig:extinct-fid}
\end{figure*} 

The lower-altitude circulation features identified in
the fiducial simulation, namely a substellar anticyclone,
an antistellar cyclone, and vertically sheared
retrograde-to-prograde transition, can be compared to
the outputs of published GCMs for WASP-121b.
\citet{2024A&A...692A.230C} found a predominantly
retrograde equatorial jet at pressures of
$\sim 1$~mbar in their GCM, qualitatively consistent
with our lower-altitude retrograde flow.
However, GCMs typically predict a deep eastward
equatorial jet at pressures $> 0.1$~bar
\citep[e.g.,][]{2019ApJ...886...26T},
whereas our reflecting inner boundary condition
($p_{\rm in}=0.83$~bar) suppresses deep convective
forcing and may suppress the prograde jet component
at the deepest modeled layers.
The vertical shear direction (retrograde at lower
altitudes transitioning to day-to-night flow at higher
altitudes) is broadly consistent with the circulation
patterns inferred from phase-resolved spectroscopy
\citep{2025Natur.639..902S}.
We emphasize that the Kratos simulations are designed
to resolve the upper atmosphere ($p \lesssim
10^{-3}$~bar) where GCMs approach their limit of
validity; the lower-boundary circulation should be
interpreted as a dynamically self-consistent but
approximate lower boundary condition for the outflow
region rather than a full GCM-equivalent treatment.

\subsubsection{Higher altitudes and outflows}

At higher altitudes, the atmospheric dynamics is dominated
by outflows. These energetic outflows, driven primarily by
XUV irradiation and assisted by optical radiation (see also
the analyses in \S\ref{sec:results-para}) from the host
star, eventually organize into the large-scale spiral
structures that characterize the extended exosphere.  This
escaping material organizes into distinct spiral-shaped
structures, a morphology reminiscent of (but with gas flow
opposite to) the accretion streams observed in the
interactions between planets and their residing
protoplanetary disks \citep[e.g.,][]{2012ARA&A..50..211K}.

The formation of the two relatively dense spiral arms is
dominated by the interaction of the planetary outflow with
the stellar gravitational field and the Coriolis force in
the co-rotating frame. On the dayside hemisphere, materials
could be heated to spill over the L1 point and accelerated
towards the star. Simultaneously, gas flowing from the
evening terminator (the boundary between day and night) is
deflected by the Coriolis force towards the star, which
could also be understood as the local sub-Keplerian motion
caused by the planetary orbital speed plus the outflow
velocity (opposite to the orbital motion direction). These
two sources together shape a dense, leading spiral arm that
points toward the host star and with an angular velocity
slightly exceeding that of the planet itself. Conversely, on
the nightside hemisphere, a trailing arm is formed by
analogous processes near the L2 point and the morning
terminator. This arm lags behind the planet's orbital
motion. The resulting bimodal spiral structure, illustrated
in Figure~\ref{fig:scheme}, dominates the large-scale
morphology of the extended exosphere.

On the bases of these dense
($\rho \sim 10^{11}~m_p~\cm^{-3}$) and relatively cool
(close to the equilibrium temperature;
$T\sim 3\times 10^3~\K$) spiral rams as ``spines'', the
incident EUV photons further trigger secondary expansion,
forming extended arms that are predominantly ionized and
heated to $T \sim 3 \times 10^4~\K$, and attenuated to
$\rho \sim 10^8~m_p~\cm^{-3}$. Both the dense and the
extended spiral arms are subject to the Coriolis force,
inherent to the co-rotating frame of reference, establishes
this spiral pattern. Material in the leading arm (advancing
ahead of the orbital motion) is deflected outward, resulting
in a net redshift in the planetary rest frame when detected,
while material in the trailing arm produces a corresponding
blueshift. We note that the planetary spin, assumed to be
tidally locked, does not exert a significant influence on
the large-scale outflow dynamics at altitudes several Earth
radii above the inner boundary, at least in the absence of
strong magnetic coupling.

As indicated by the velocity streamlines, the dense
spiral-arm features originate mainly by in the vicinity of the
L1 and L2 Lagrangian points, when the heated atmosphere at lower
altitudes fills up the Roche lobe, spilling out, and guided by
the morphologies of equipotential surfaces. The gas inside dense 
spiral arms undergoes acceleration from near-stagnation conditions
at the base to transonic velocities near the Lagrangian
points. The divergent streamlines of the flow above these
points form nozzle-like conditions that further accelerate
the outflow to supersonic velocities, enabling escape from
the planet gravitational potential.

\subsection{Thermochemical Conditions and Detectables}
\label{sec:thermochem-spec}

Observational insights into the kinematics and
thermochemistry are obtained by linking synthetic spectra to
the underlying thermochemical species distributions in the
simulation. As illustrated in Figures~\ref{fig:slice-fid}
and \ref{fig:render-fid}, the four primary spectroscopic
tracers (neutral sodium Na, neutral iron Fe, H$\alpha$, and
metastable helium \ext{He}) exhibit distinct spatial
distributions that reflect their differing population
mechanisms and vulnerability to photoionization and
collisional ionization. Note that neutral iron has numerous
absorption lines, and we select one representative line at
$\lambda = 4490~\ang$ for kinematic analysis, which
adequately captures the gas motions revealed by iron
features. For sodium, the two D lines are separated by
$\sim 6~\ang$ (equivalent to $\sim 300~\km~\s^{-1}$); we
consider only the $5896~\ang$ line as representative.

\begin{figure*}
  \centering\includegraphics[width=0.32\linewidth]
  {\figdir/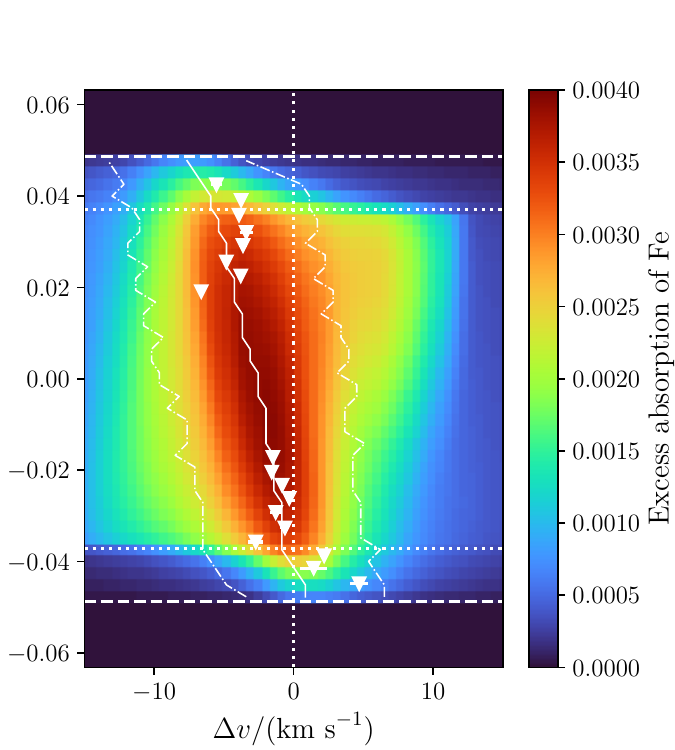}  
  \includegraphics[width=0.32\linewidth]
  {\figdir/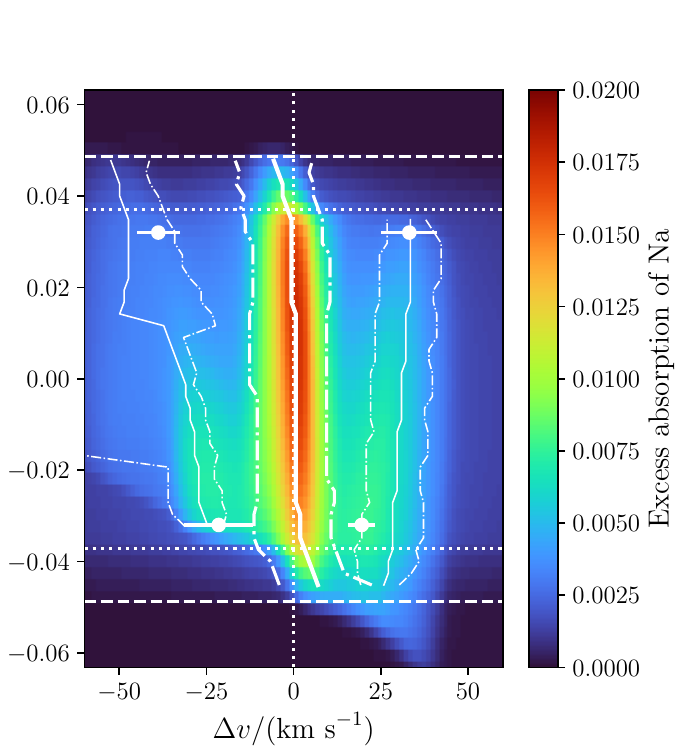}
  \includegraphics[width=0.32\linewidth]
  {\figdir/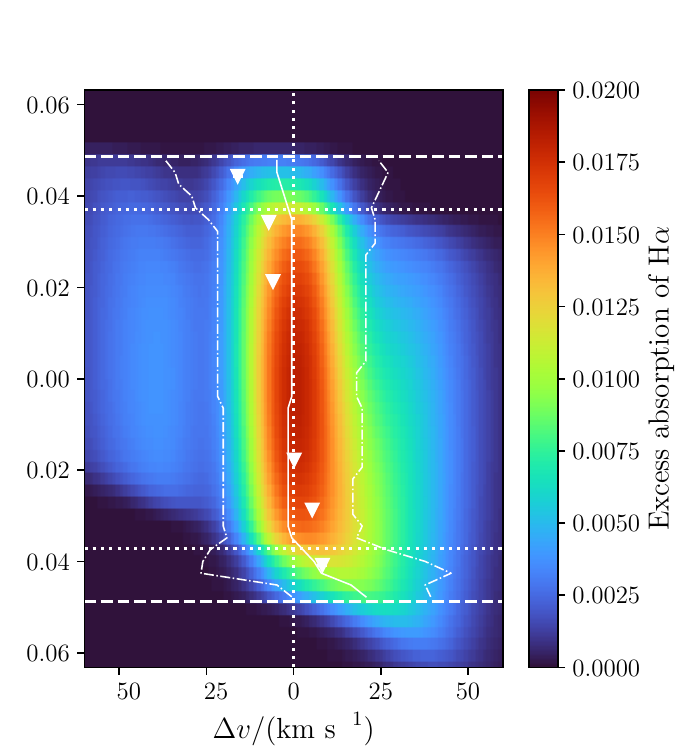}
  \caption{Absorption spectra of Fe (left panel), Na (middle
    panel), and H$\alpha$ (right panel), illustrated as
    excess absorption at different velocities (horizontal
    axes) and orbital phases (vertical axes). Horizontal
    white dashed lines indicate the starts of ingress and
    the ends of egress, and horizontal white dotted lines
    indicate the ends of ingress and starts of egress. The
    vertical dotted lines indicate zero velocity shifts for
    reference. The absorption peaks, evolving with orbital
    phases, are tracked with solid lines for the centroids,
    and the dash-dotted lines for the half-maximum widths.
    For the Na panel, the central peak are tracked by heavy
    solid and dash-dotted lines. Errorbars for Fe and
    H$\alpha$ indicate the fitting results and uncertainties
    (not widths) reported by \citet{2025Natur.639..902S}. More detailed comparison of H$\alpha$ spectra is shown in Fig. \ref{fig:compare-halpha-fid} and discussed in \S \ref{sec:H_alpha}.  
    The errorbars for the Na panel indicate the
    fitting centroids
    and widths within this work (by using the observation transmission data
    presented in \citealt{2025Natur.639..902S}). More detailed comparisons of Na spectra are shown in Fig. \ref{fig:compare-na-fid} and Fig. \ref{fig:compare-na-var}. These results are discussed in \S \ref{sec:Na} and \S \ref{sec:results-para}. Observation errorbars are
    shifted to the red side by $3~\km~\s^{-1}$ for better
    fitting of the trend (note that the observed velocity
    shifts may not have a fixed calibration). }
  \label{fig:spec-fid}
\end{figure*}

Our simulations identify four principal reservoirs where
these species can survive the intense stellar irradiation:
\begin{enumerate}
\item The high-density interior region: Inside the planetary
  photosphere, extreme column densities shield neutral
  species from high-energy photons.
\item The planet shadow: The region directly behind the
  planet receives no direct stellar illumination,
  maintaining relatively cool, neutral conditions.
\item The dense spiral arms: Gas densities reaching
  $\rho \gtrsim 10^{11}~m_p~{\rm cm}^{-3}$ create optically
  thick barriers that attenuate EUV radiation, while
  allowing deeper-penetrating X-ray and FUV photons to reach
  regions closer to the planetary surface.
\item The extended spiral arm structures: The diffuse outer
  parts of the arms, heated by EUV and soft X-ray
  irradiation. The ionized fraction here is higher than in
  the first three reservoirs but still not fully
  ionized. This region favors neutral species that require
  some ionization for excitation (e.g., \ext{He}).
\end{enumerate}
These distinct reservoirs create chemically stratified
layers with different observational consequences. The
synthetic transmission spectra illustrated in
Figure~\ref{fig:spec-fid} exhibit reasonable, qualitative to
semi-quantitative agreement of velocity shifts and
amplitudes (due to the uncertainties of the actual
abundances of metals near the planet photosphere) with
\citet{2025Natur.639..902S}. The observable signatures of
different chemical species indicate different physical
mechanisms and processes.

\begin{figure*}
  \centering
  \vspace{-0.5cm} 
  \hspace{-0.8cm}\includegraphics[width=0.33\linewidth]
  {\figdir/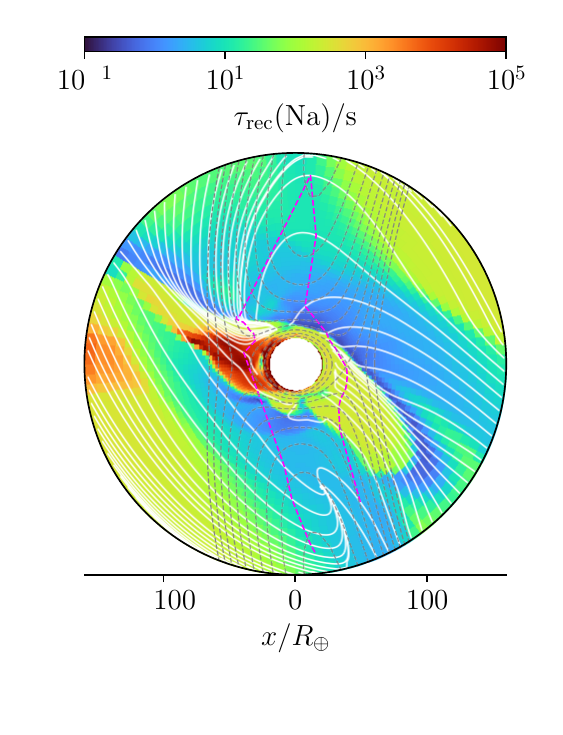}
  \hspace{-0.8cm}\includegraphics[width=0.33\linewidth]
  {\figdir/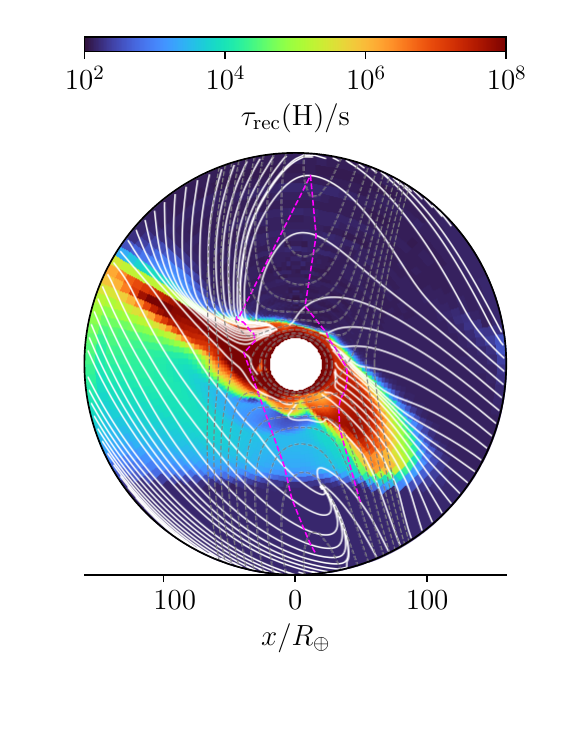}  
  \hspace{-0.8cm}\includegraphics[width=0.33\linewidth]
  {\figdir/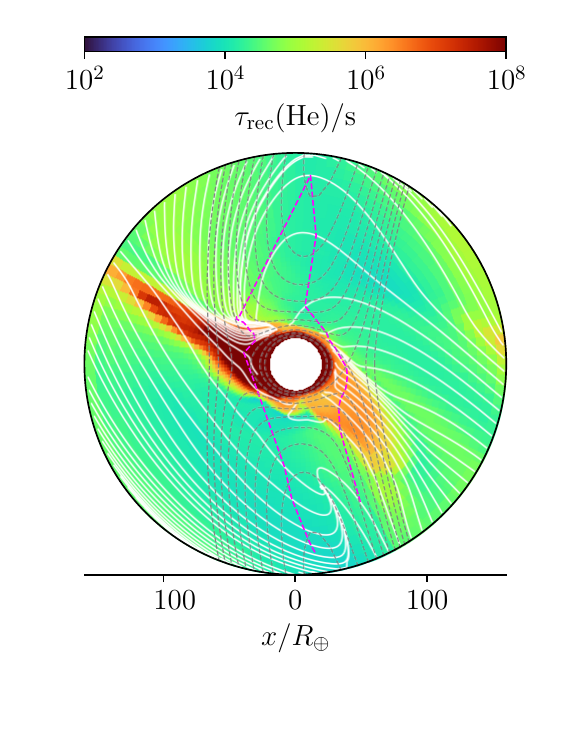}  
  \vspace{-1.1cm}
  \caption{Recombination timescales of \chem{Na} (left
    panel), \chem{H} (middle panel, to the $2s$ level, from
    \chem{H^+} via the recombination with $e^-$ and charge
    exchange with neutral H), and \chem{He} (right panel; to
    the metastable state, via the recombination from
    \chem{He^+}). }
  \label{fig:slice-recomb-fid}
\end{figure*}

\subsubsection{Fe and inner regions}

As one can confirm by inspecting
Figures~\ref{fig:render-fid} and \ref{fig:extinct-fid}, Fe
primarily tracks the kinematics of the inner layer,
primarily revealing the motion from dayside to the
nightside. When superposed with the rotation of the frame
(originating from the tidally locked spin), whose linear
speed satisfies,
\begin{equation}
  \label{eq:planet-spin}
  \begin{split}
    v_{\rm rot}
    & = \left(\dfrac{G M_*}{a^3}\right)^{1/2} R_{\rm p} \\
    & = 6.9~\km~\s^{-1} \times
      \left(\dfrac{a}{0.026~\au}\right)^{-3/2}
      \left(\dfrac{R_{\rm p}}{20 R_\oplus}\right)^{-3/2},
  \end{split}
\end{equation}
the velocity signal identified by Fe becomes slightly
redshifted at ingress and systematically blueshifted
throughout the rest of the transit. In the time-dependent
transmission spectra, Fe absorption profiles exhibit the
characteristic tidally locked spin signature (redshift at
ingress, blueshift at egress), superposed with a systematic
blueshift of $\sim 3.5~\km~\s^{-1}$ that traces day-night
circulation patterns. This is in quantitative agreement with
the results of \citet{2025Natur.639..902S}, comparing the
observation errorbars and the excess absorption colormap
illustrated in Figure~\ref{fig:spec-fid}.

\begin{figure} 
  \centering
  \includegraphics[width=0.99\linewidth]
  {\figdir/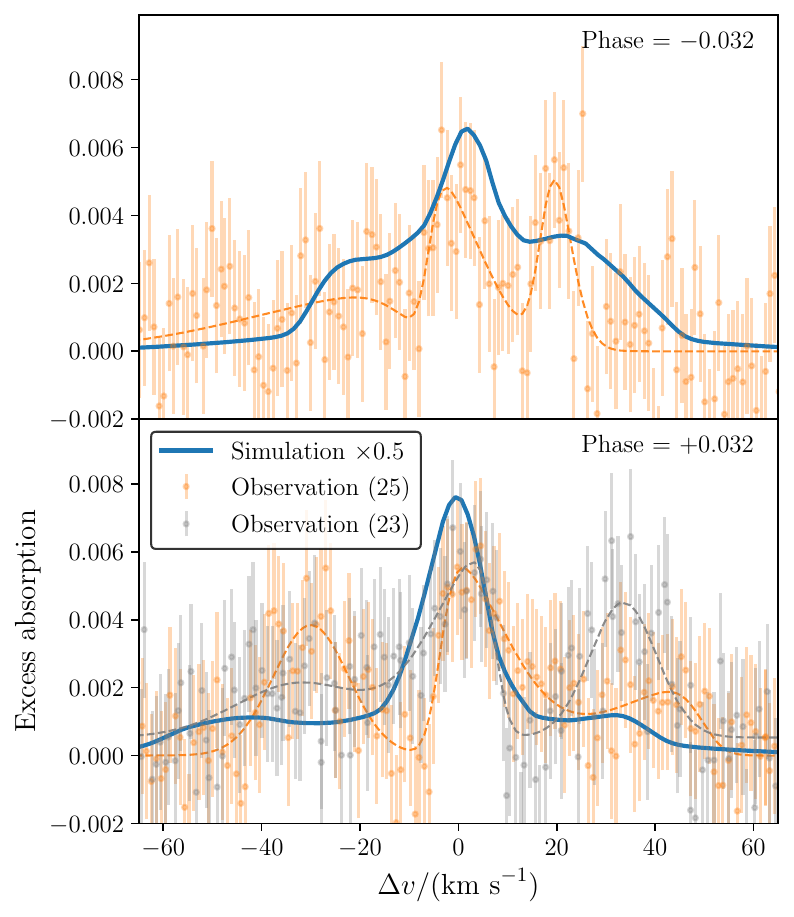} 
  \caption{Comparisons of excess absorption of Na, between
    the fiducial simulation results (in solid blue lines;
    multiplied by a factor of $0.5$, see
    \S\ref{sec:thermochem-spec}), the observations indicated
    with errorbars in \citet{2023A&A...673A.125S} [grey
    errorbars, egress only, marked as ``Observation (23)'']
    and \citet{2025Natur.639..902S} [orange errorbars, both
    ingress and egress, marked as ``Observation (25)''], and
    the fitting results within this work (using the errorbar
    data, presented in dashed lines in the same color as the
    corresponding errorbars). }
  \label{fig:compare-na-fid}
\end{figure}

\begin{figure*}
  \centering
  \includegraphics[width=0.9\linewidth]
  {\figdir/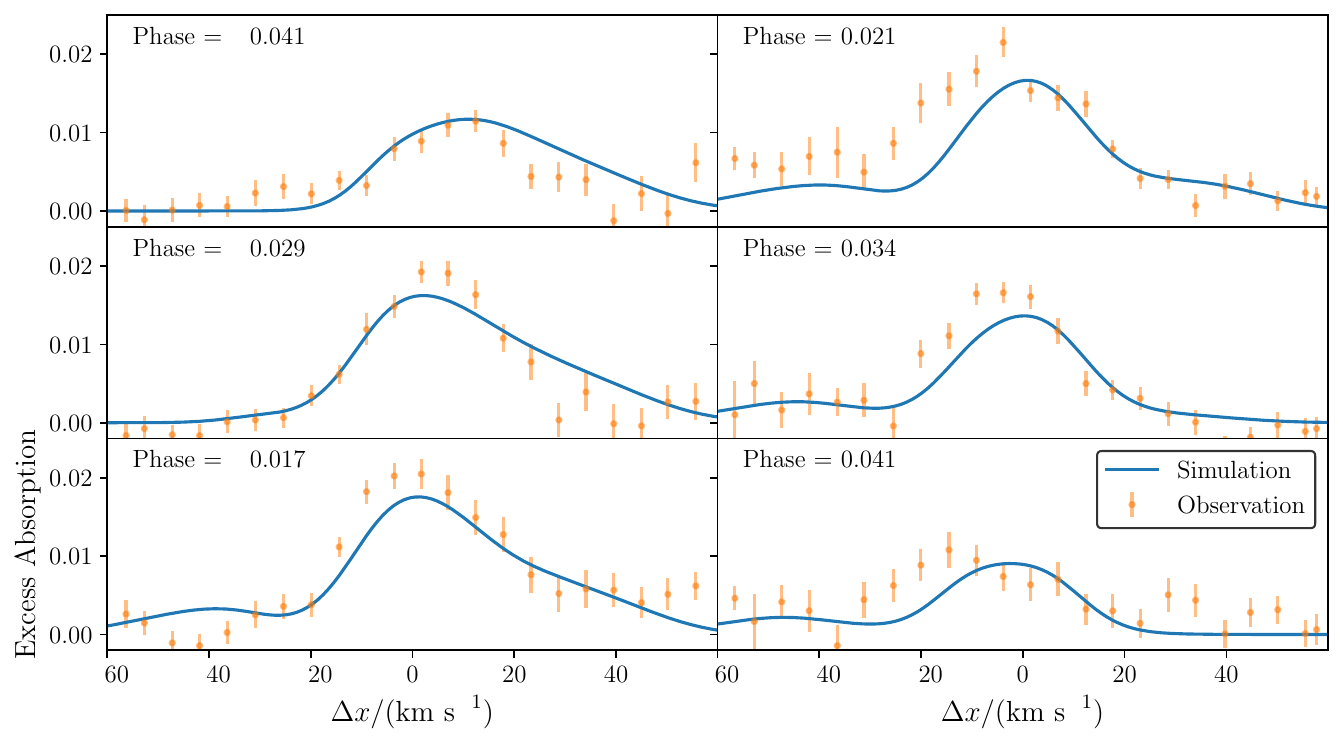}  
  \caption{Comparing the simulated transmission spectra of
    H$\alpha$ to the observations (oragne errorbars; from
    \citealt{2025Natur.639..902S}) at different orbital
    phases, near the ingress (left column) and egress (right
    column), respectively. Note the secondary peaks,
    redshifted at Phase$<0$ (near ingress), and blueshifted
    at Phase$>0$ (near egress).  }
  \label{fig:compare-halpha-fid}
\end{figure*}

\subsubsection{Na and the dense spiral arms}
\label{sec:Na}

At altitudes higher than where neutral Fe survives,
velocities are also influenced by orbital motion. The
spatial distribution of Na that contributes to the
transmission spectra can be divided into the dense
reservoirs near the planet and the spiral arms, the latter
of which exhibits relatively large velocities (see also
Figure~\ref{fig:extinct-fid}). One prospective curiosity is
how neutral Na survives the intense FUV radiation
(especially in the $6~\eV$ band with intense photon flux),
which can penetrate the dense spiral arms almost
unattenuated. At the peak band intensity
$F(6~\eV) = 2\times 10^{18}~\cm^{-2}~\s^{-1}$, the
photoionization rate of Na is
$\zeta(\chem{Na}) \simeq 3\times
10^{-3}~\s^{-1}$. Meanwhile, the recombination timescale in
the spiral arms is approximately
\begin{equation}
  \label{eq:na-recomb}
  \begin{split}
  \tau_{\rm rec}(\chem{Na})
    & \equiv \dfrac{n(\chem{Na})}
    {\alpha_{\chem{Na}} n(\chem{Na^+})n_e}
      \\
    & \simeq 3\times 10^2~\s\ \times
  \left(\dfrac{T}{10^4~\K}\right)^{0.68}
      \left(\dfrac{n_e}{10^8~\cm^{-3}}\right)^{-1}
    \\
    & \quad \times
  \left[\dfrac{n(\chem{Na^+})}{10^4~\cm^{-3}}\right]^{-1}
  \left[\dfrac{n(\chem{Na})}{10^2~\cm^{-3}}\right],
  \end{split}
\end{equation}
using the recombination rate from \citet{2013A&A...550A..36M} (also
implemented in our simulations). As
$\tau_{\rm rec}(\chem{Na})\sim 1/\zeta(\chem{Na})$,
recombination can offset ionization and maintain a
significant neutral sodium fraction in the spiral arms (see
also Figure~\ref{fig:slice-recomb-fid}).  The velocity
magnitudes within the spirals, estimated from stellar
potential acceleration $\Delta v_{\rm r}$ and Coriolis
deflection $\Delta v_{\rm co}$ (where $\Delta r$ is the
distance traveled after leaving the Roche lobe),
\begin{equation}
  \label{eq:vel-arms}
  \begin{split}
    & |\Delta v_{\rm r}| \sim
    \left( \dfrac{GM_*}{2 a^2}\Delta r\right)^{1/2}
    \sim 40~\km~\s^{-1}\times \left( \dfrac{\Delta r}
      {2~R_{\rm p}} \right)^{1/2}\ , \\
    & |\Delta v_{\rm co}| \sim 2 \Omega \Delta r \sim
      30~\km~\s^{-1} \times  \left( \dfrac{\Delta
      r}{2~R_{\rm p}} \right)\ ,
  \end{split} 
\end{equation}
are of order $\sim 30~\km~\s^{-1}$ at $\sim 2~R_{\rm p}$
from the planetary surface.  Although these velocities could
increase at larger distances, the observed velocity shifts
in secondary Na excess absorption are limited to
$|\Delta v| \lesssim 50~\km~\s^{-1}$ \citep[see
also][]{2025Natur.639..902S}. This observational constraint
implies that the Na-rich spiral arm material must be
confined to regions relatively close to the planet,
typically within $\lesssim 5~R_{\rm p}$. Our simulations
indicate that this confinement is enforced by the intense
FUV flux from the F6V host star WASP-121, which limits the
survival of neutral sodium to these inner regions.
Geometric effects during the transit further modulate the
observed velocity shifts. During ingress, the leading spiral
arm presents a larger spatial extent but a smaller LoS
velocity component because its motion is primarily
perpendicular to the LoS. Conversely, during egress, the
trailing arm presents a more favorable geometry for
observing Doppler shifts (see Figure~\ref{fig:scheme}). As a
result, the velocity shift immediately after ingress is
$\Delta v \simeq 20~\km~\s^{-1}$, considerably smaller than
the blueshift magnitude on the egress side
($|\Delta v|\simeq 35~\km~\s^{-1}$).

Comparison between our simulated transmission spectra and
observations from the morning (ingress) and evening (egress)
limbs (Fig.~\ref{fig:compare-na-fid}) shows
semi-quantitative agreement, though the model amplitudes
have been scaled down by a factor of $0.5$ to match the
data. This scaling is consistent with uncertainties in the
planetary sodium abundance, which is known to be substellar
for WASP-121b \citep{2023A&A...673A.125S}.

A detailed inspection of the observed transmission spectra
reveals intriguing asymmetries not fully captured by
previous atmospheric models. During ingress, a relatively
obscured but marginally detectable blueshifted absorption
component appears alongside the dominant redshifted
secondary peak. Conversely, during egress, a noticeable
redshifted tertiary feature coexists with the primary
blueshifted absorption. 

These ``tertiary'' peaks, not fully
reproduced in existing spectral retrievals
\citep{2025Natur.639..902S}, are difficult to reconcile with
a uniform prograde super-rotating jet, which would generate
largely symmetric morning and evening signatures.

Moreover, the velocity separation between the secondary and
tertiary absorption features differs between the morning and
evening limbs, indicating a distinct morning-evening
asymmetry in atmospheric kinematics. The 3D spiral-arm model
could accounts for these complex spectral features. While
the leading spiral arm dominates absorption on the morning
limb, the trailing arm also contributes with oppositely
directed velocity shifts due to Coriolis
deflection. Additionally, geometric projection effects cause
the spiral arms to appear more spatially extended but less
velocity-broadened during ingress (when they are more
perpendicular to the line of sight), with the reverse
occurring during egress (see Fig.~\ref{fig:scheme}).  We
also notice that the amplitudes and velocity separations of
the secondary and tertiary peaks show variability between
the observational datasets of \citet{2023A&A...673A.125S}
and \citet{2025Natur.639..902S}. Such variability is consistent with
intrinsic hydrodynamic instabilities in the atmospheric
outflow (see the KHI discussion in \S\ref{sec:morph-advection}). Therefore, our spiral-arm
model provides a complimentary explanation for the observed Na
line profiles without requiring extreme near-surface jet
streams with velocities comparable to the planetary escape
velocity (equation~\ref{eq:v-esc}).

\subsubsection{\chem{H}$\alpha$ and metastable helium in the extended arms}
\label{sec:H_alpha}

\begin{figure}
  \centering
  \includegraphics[width=0.8\linewidth]
  {\figdir/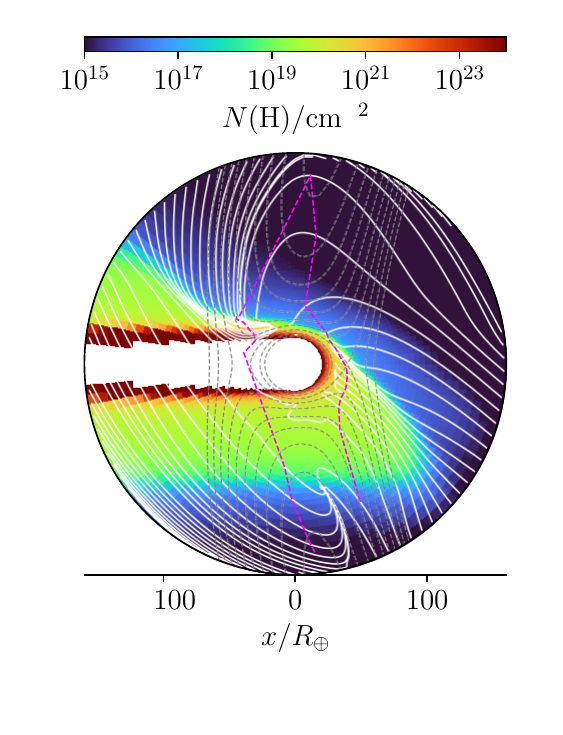}
  \vspace{-0.8cm}  
  \caption{Slice plot in the equitorial plane (similar to
    Figure~\ref{fig:slice-fid}), showing the column density
    of neutral atomic hydrogen (H) calculated from the
    ray sources. }
  \label{fig:slice-ncol-h}
\end{figure}

The H$\alpha$ transmission spectral profiles, as shown in
Figures~\ref{fig:slice-fid} and \ref{fig:render-fid},
originate from extended regions predominantly located
outside the planetary Roche lobe. These regions are
dynamically coupled to orbital motion and stellar gravity
rather than to the planet spin.  The collisional excitation
timescale to the $2s$ state (the excitation to the $2p$
state would instead lead to instant \lya emission) is
approximately,
\begin{equation}
  \label{eq:ha-coll-pop}
  \begin{split}
    \tau_{\rm coll}(\chem{H}^{2s})
    & \equiv \dfrac{1}{k_{1s\rightarrow 2s} n_e} 
    \\
    & \sim 1~\s\times \exp
      \left(\dfrac{118400~\K}{T}\right)
      \left(\dfrac{T}{10^4~\K}\right)^{-0.455}\ ,
  \end{split}
\end{equation}
with the collisional excitation rate coefficient
$k_{1s\rightarrow 2s}$ taken from the
\citet{2006agna.book.....O} compilation.
This timescale increases to
$\tau_{\rm coll}\sim 10^5~\s$ at $T = 10^4~\K$ near the
interface of the dense to extended spiral arms, and to
$\tau_{\rm coll}\sim 2\times 10^{10}~\s$ at $T = 5000~\K$
within the denser arms, making recombination and charge
exchange the dominant population mechanisms for
the non-\lya channels (middle panel of
Figure~\ref{fig:slice-recomb-fid}). Consequently, H$^{2s}$
(the $2s$ state of neutral hydrogen responsible for
H$\alpha$ absorption) resides in regions with a relatively
high ionization fraction, specifically tracing the interface
between the dense spiral arms and the EUV-irradiated
extended arms where hydrogen is nearly fully ionized.

Comparison of the simulated transmission spectra with
observations (Figure~\ref{fig:spec-fid}) shows that the
orbital phase dependence of the peak velocity generally
matches the observed trend. Notably, both the simulations
and the data reveal secondary absorption peaks at velocities
$|\Delta v| \gtrsim 40~\km~\s^{-1}$ for H$\alpha$, features
that were not explicitly identified by
\citet{2025Natur.639..902S}. Similar to neutral Na, these
secondary peaks arise because the interface between the
dense and extended spiral arms extends to distances
$\gtrsim 10^2~R_\oplus$ from the planet, where gas is
accelerated to $\sim 40~\km~\s^{-1}$ by the stellar
gravitational potential and Coriolis force
(eq.~\ref{eq:vel-arms}).
Figure~\ref{fig:compare-halpha-fid} demonstrates
  good agreement in both the central and secondary
  high-velocity peaks across multiple orbital phases (for
  reference, the H$\alpha$ spectra computed with the
  alternative scheme are provided in
  Appendix~\ref{sec:appendix-ha-old}).  These secondary
peaks could be overlooked in previous analyses that did not
account for spiral arm structures. 

Similarly, metastable helium (\ext{He}) absorption arises in
regions where recombination populates the metastable triplet
ground state of helium (right panel of
Figure~\ref{fig:slice-recomb-fid}). The formation
and destruction pathways for He($2^3S$) are described in
\S\ref{sec:method-spec}. A test simulation including
explicit He$^{2+}$ species confirms that while He$^{2+}$
dominates over He$^+$ in the diffuse outer regions of the
domain, the contribution of those regions to the
He~10830~\ang\ optical depth is negligible ($<10\%$
change in the equivalent width;
Appendix~\ref{sec:appendix-hepp}).
The \ext{He} absorbers
are spatially more extended than those of H$\alpha$, though
their number density is considerably lower. Since \ext{He}
is efficiently destroyed by photoionization from photons
above $4.8~\eV$, a prominent shadow tail forms behind the
planet (see Figure~\ref{fig:render-fid}), a feature
previously identified in studies of evaporating exoplanets
\citep[e.g.,][]{2021ApJ...914...98W,2021ApJ...914...99W}. Due
to the finite extent of our simulation domain, we cannot
compare the equivalent width of the He $10830~\ang$ line
over the entire orbital period with observations. At least,
the dimensionless equivalent width in the
$10830 < (\lambda/\ang)< 10840$ band defined as ($F_{\rm c}$
for the unabsorbed contiuum component of the spectrum, and
$F_\lambda$ for the spectrum with absorption;
$\Delta \lambda = 10~\ang$),
\begin{equation}
  \label{eq:def-equiv-width}
  [W/\Delta \lambda](\chem{He}) \equiv
  \dfrac{1}{\Delta \lambda} \int_{10830~\ang}^{10830~\ang+
    \Delta\lambda} \d\lambda\, \dfrac{F_{\rm c} - F_\lambda
  } {F_{\rm c} \Delta \lambda}\ ,
\end{equation}
is obtained from our simulation
$[W/\Delta\lambda](\chem{He})\sim 3600~{\rm ppm}$ at
mid-transit, agrees quantitatively with the values reported
by \citet{2024A&A...692A.230C, 2025NatCo..1610822A} when
accounting for the fact that the observed in-transit
absorption is an average over the full transit, which tends
to yield a shallower absorption depth
($\sim 2800~{\rm ppm}$) than the mid-transit value.

\section{Parametric Study}
\label{sec:results-para}

While the fiducial model provides a baseline understanding,
the interactions of physical processes governing atmospheric
escape necessitates a broader exploration of the parameter
space. To systematically quantify the sensitivity of our
results and identify the dominant physics shaping the
observables, we conducted a series of controlled numerical
experiments summarized in Table~\ref{table:models-var},
mainly experimenting the impacts of the stellar radiation
and winds. These simulations are also illustrated in
Figures~\ref{fig:multimodel} and \ref{fig:multimodel-cont},
showing especially the profiles of Na which traces the most
prominent spiral arm features.

\renewcommand{\arraystretch}{1.2}
\begin{deluxetable*}{llll}
  \tablecolumns{6} 
  \tabletypesize{\scriptsize}
  \tablewidth{0pt}
  \tablecaption{Various models based on 
    the fiducial model for WASP-121b 
    \label{table:models-var}
  } \tablehead{ \colhead{Model Code} & \colhead{Description}
    & \colhead{$\dot{M}/(10^{-7}~M_\oplus~\yr^{-1})$} &
    \colhead{$[W/\Delta \lambda]({\rm He})^\dagger/10^{-3}$} }
  \startdata 0 & Fiducial Model (\S\ref{sec:results})
  & 0.64 & 3.6 \\
  OPT3 & $3\times$ flux at $h\nu = 2~\eV$.
  & 0.75 & 3.5 \\
  FUV10 & $10\times$ fluxes at
  $(h\nu/\eV) \in \{4.9, 6, 12\}$
  & 11.8 & 1.5 \\
  EUV10 & $10\times$ fluxes at $(h\nu/\eV) \in \{20, 60\}$
  & 1.17 & 23.9 \\
  XR10 & $10\times$ fluxes at $(h\nu/\keV) \in \{0.3, 3 \}$
  & 4.5 & 18.4 \\
  DUST & $n({\rm Gr})/n_{\rm H} = 10^{-9}$
  & 0.96 & 3.3 \\
  WIND &
  $\dot{M}_{\rm wind} = 6\times 10^{-12}~M_\odot~{\rm
    yr}^{-1}$ & 0.26 & 10.9 \enddata
  \tablecomments{$\dagger$: $[W/\Delta \lambda]$ is the
    dimensionless equivalent width of He $10830~\ang$
    absorption line in the wavelength band
    $10830 < (\lambda/\ang) < 10840$ (equivalent to the
    absorbed fraction in the same band) at the
    mid-transit (see eq.~\ref{eq:def-equiv-width}). }
\end{deluxetable*}

\subsection{Dependence on High-Energy Radiation Flux}
\label{sec:results-var-rad}

The spectral energy distribution of the host star plays a
dual role, by providing the energy that drives atmospheric
escape, but also ionizes and dissociates the very atomic and
molecular species used to trace the outflow. Our simulations
examine how different energy bands distinctly shape the
observable properties of the escaping atmosphere, with each
band influencing the outflow density, velocity, and chemical
composition in different ways.

Model FUV10 (which has an FUV flux enhanced by a factor of
10 relative to the Model 0) demonstrates that FUV radiation
primarily controls the abundance and distribution of tracer
species via photoionization and launching outflows. FUV
photons with energies above $4.8~\eV$ can photoionize
metastable helium, while those above $5.14~\eV$ ionize
neutral sodium.  Regarding the absence of a
  dedicated \lya (10.2~eV) radiation bin, we note that
  the 12~eV (representing $7.9$ to $13.6~\eV$) already
  included the effective photochemical effects of \lya
  photons. The photon flux in this bin,
  $10^{16}~\cm^{-2}~\s^{-1}$ at $0.026~\au$, corresponds to
  an equivalent \lya luminosity $\sim 20\times$ the
  solar \lya luminosity
  \citep{2000JGR...10527195W}. In comparison, the host star
  WASP-121, an F6V star, is expected to have a \lya
  luminosity of a few to $\sim 10\times$ solar; (e.g.,
  \citealt{2013ApJ...766...69L}).  Even accounting for
  resonant amplification by the factor of a few in high
  optical depth as found by \citet{2017ApJ...851..150H}, the
  energy budget is covered with margin.

\begin{figure*}
  \centering
  \raggedright OPT3 \\ \vspace{-0.2cm} \hspace{-1.0cm}
  \includegraphics[width=0.33\linewidth,valign=t]
  {\figdir/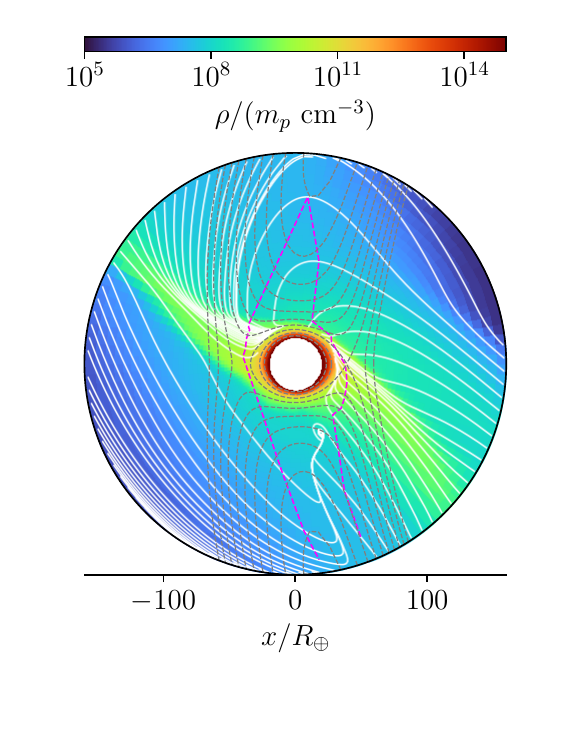}
  \hspace{-0.8cm}
  \includegraphics[width=0.33\linewidth,valign=t]
  {\figdir/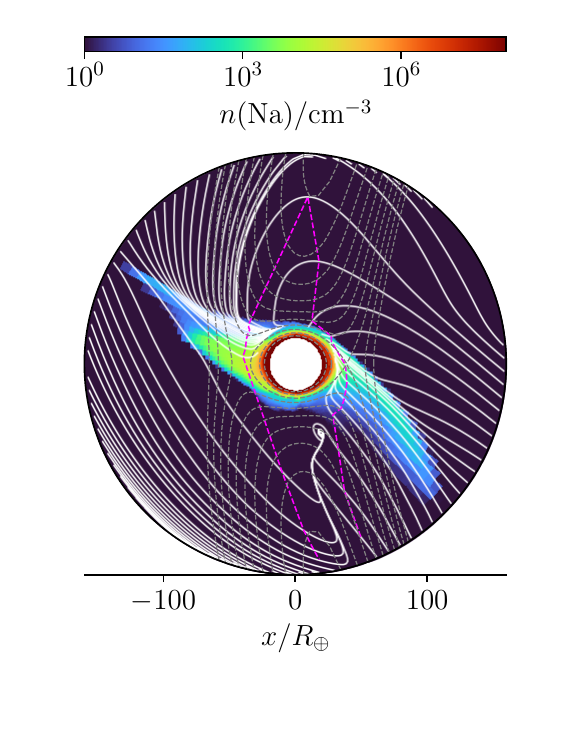}
  \includegraphics[width=0.32\linewidth,valign=t]
  {\figdir/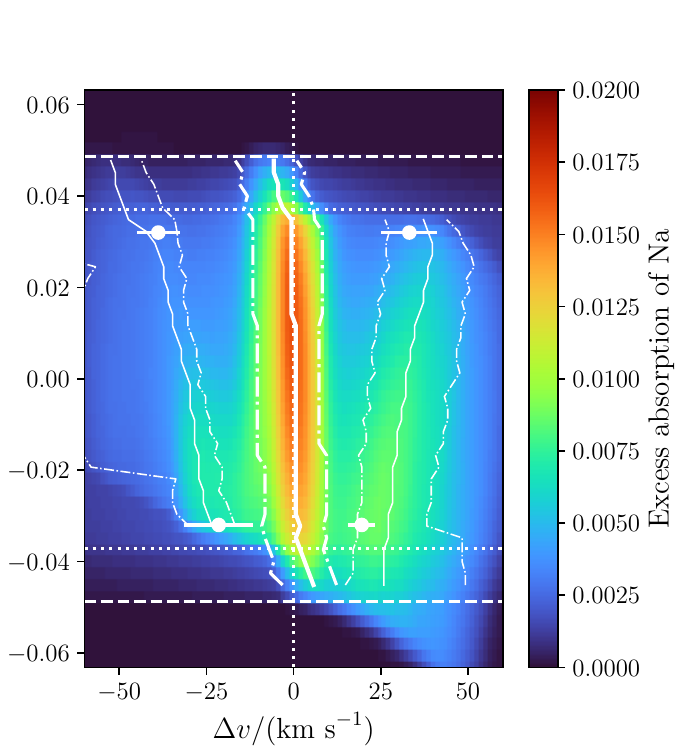}  
  \\ \vspace{-0.8cm}
  \raggedright FUV10 \\ \vspace{-0.2cm} \hspace{-1.0cm}
  \includegraphics[width=0.33\linewidth,valign=t]
  {\figdir/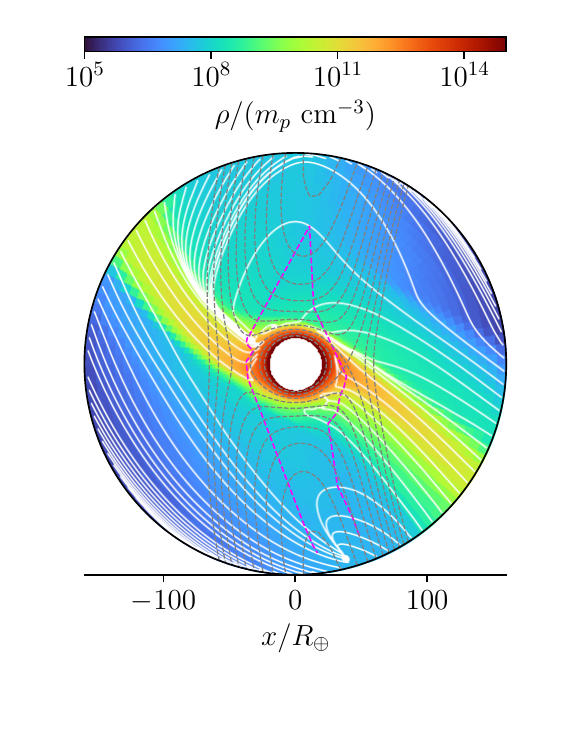}
  \hspace{-0.8cm}
  \includegraphics[width=0.33\linewidth,valign=t]
  {\figdir/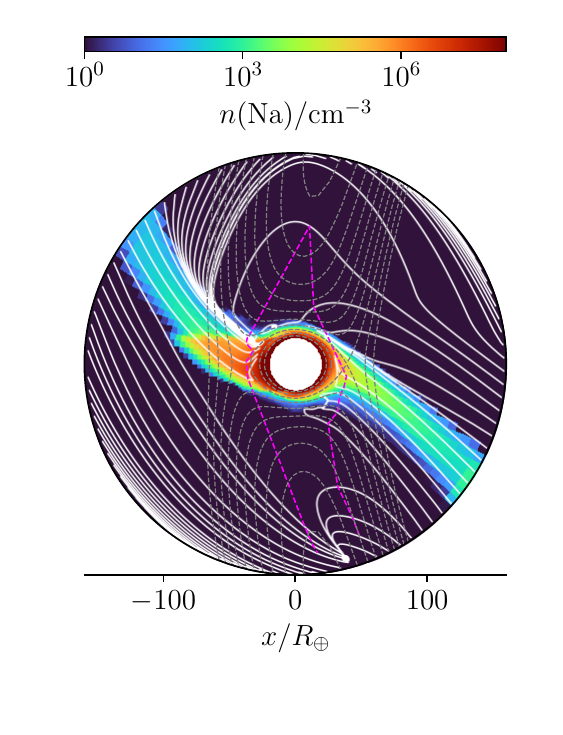}
  \includegraphics[width=0.34\linewidth,valign=t]
  {\figdir/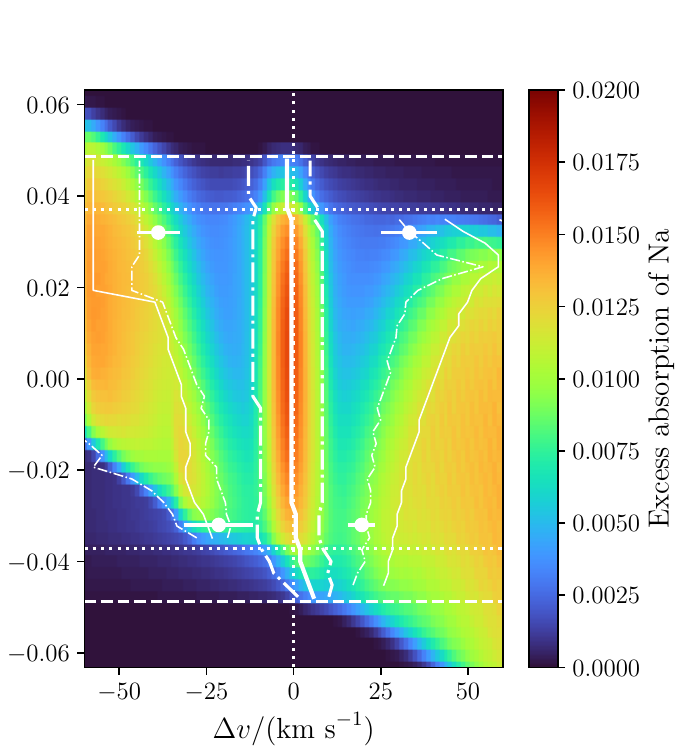}
  \\ \vspace{-0.8cm}
  \raggedright EUV10 \\ \vspace{-0.2cm} \hspace{-1.0cm}  
  \includegraphics[width=0.33\linewidth,valign=t]
  {\figdir/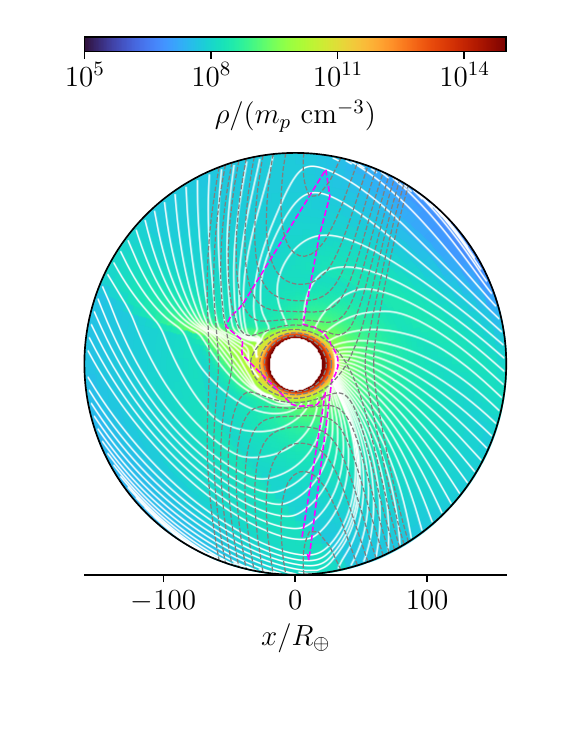}
  \hspace{-0.8cm}
  \includegraphics[width=0.33\linewidth,valign=t]
  {\figdir/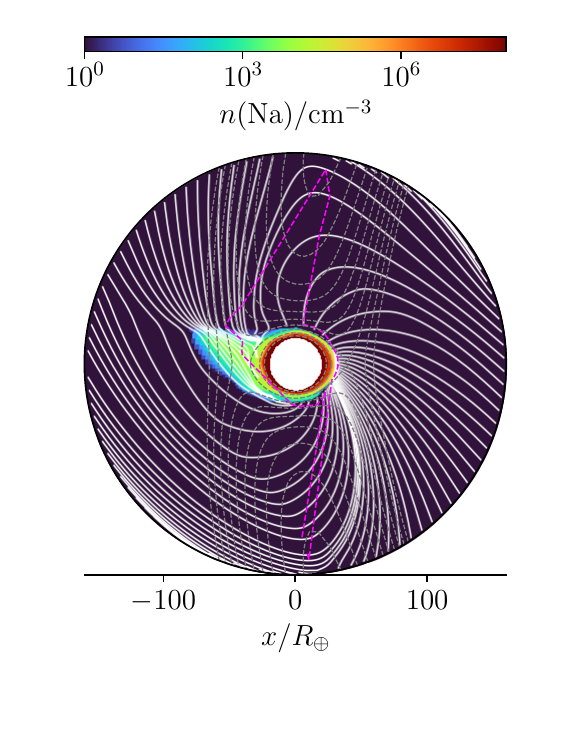}
  \includegraphics[width=0.34\linewidth,valign=t]
  {\figdir/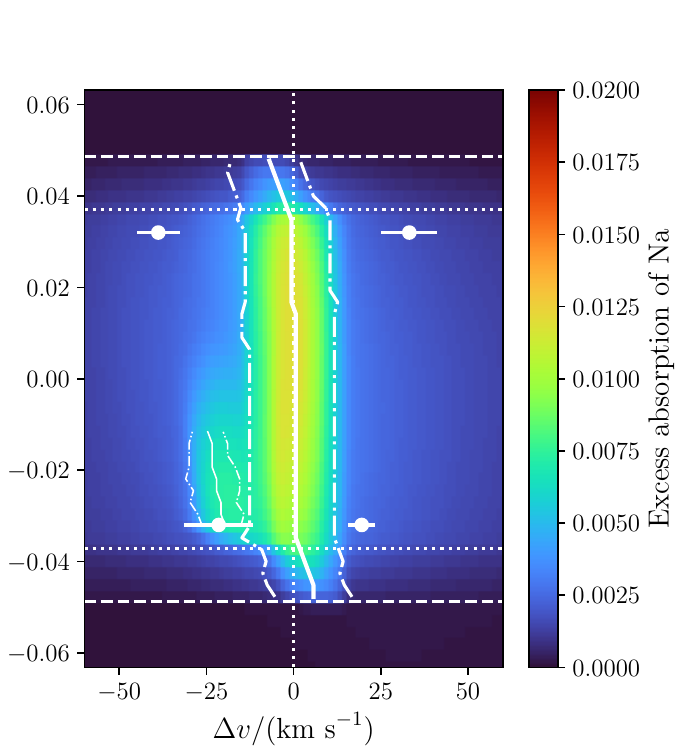}
  \vspace{-0.8cm}
  \caption{Multiple models (OPT3, FUV10, and EUV10; marked
    on the top-left of each row) exploring the responses to
    physical parameters of the planetary outflows
    (\S\ref{sec:results-para} and
    Table~\ref{table:models-var}), showing the mass density
    (left column) and neutral Na (middle column) in the
    equitorial plane (similar to
    Figure~\ref{fig:slice-fid}), as well as the excess
    absorption spectra by Na (similar to
    Figure~\ref{fig:spec-fid}). }
  \label{fig:multimodel}
\end{figure*}

In Model FUV10, the stronger FUV field leads
to a more vigorous outflow, while the resulting higher
densities in the spiral arms help maintain a relatively
abundant population of neutral Na via recombination
(although with much stronger photoionization) out to larger
distances from the planet despite the enhanced
photoionization rate (see Figure~\ref{fig:multimodel}, row
FUV10). Stronger outflows make the secondary and tertiary
peaks in the morning-evening Na transmission spectra
significantly more prominent, and move them further away
from the $\Delta v = 0$ center as they travel and get
accelerated farther from the planet (see also
Figure~\ref{fig:compare-na-var}). As the Na secondary peaks
appear to be less prominent than the observed data in the
fiducial model, it is possible that the real values of FUV
fluxes stays somewhere between Models 0 and FUV10.
Conversely, a simulation with an FUV field reduced to
one-tenth of the fiducial value (not shown in this paper)
yields much less dense spiral arms, exhibiting a narrower
spatial and velocity extent of Na absorption, as well as a
smaller equivalent width of the He $10830~\ang$ line. We
note, however, that such a weak-FUV scenario is
astrophysically unlikely for a star like WASP-121, whose FUV
flux originates primarily from the stellar blackbody
component.

In the current Model 0, which lacks solid grains (presumed
to be fully sublimated at the high temperatures of
$\sim 2400~\K$), one of the main opacity sources is
\chem{H^-}, which is susceptible to photoionization by
optical radiation (represented by the $h\nu = 2~\eV$
band). However, because the regions where \chem{H^-} absorbs
optical photons are deep within the gravitational potential
well (near the planetary surface), Model OPT3 (with optical
flux enhanced by a factor of 3) launches the outflow at a
rate only slightly stronger than the fiducial model
(approximately $1.3\times$ the mass-loss rate). This model
also produces an excess Na absorption peak on the ingress
side that extends to a redshift velocity of
$\sim 40~\km~\s^{-1}$.

The DUST model (Table~\ref{table:models-var}) is
  included intended as a bracketing upper-limit case,
  assuming the maximum plausible survival of refractory
  carbonaceous grains (graphites, or polycyclic aromatic
  hydrocarbons, PAHs, which can survive at much higher
  sublimation temperature ($\simeq 3000$~K); see also
  \citealt{2026ApJ...997...14X} and references therein) to
  test whether grain opacity affects the observable
  diagnostics.  As an additional opacity source, the assumed
  dust grains enhance the atmospheric mass-loss rate to even
  higher than OPT3.  These grains efficiently absorb
stellar radiation and transfer energy to the gas,
effectively increasing the heating efficiency in the lower
atmosphere. Both Models OPT3 and DUST exhibit synthetic Na
absorption features that appear closer to the observation
than the fiducial Model 0, in terms of secondary and
tertiary peak locations and amplitudes
(Figure~\ref{fig:compare-na-var}), yet more deterministic
conclusions could only be addressed with future observations
with higher spectral SNRs.

\begin{figure*}
  \centering
  \raggedright XR10 \\ \vspace{-0.2cm} \hspace{-1.0cm}
  \includegraphics[width=0.33\linewidth,valign=t]
  {\figdir/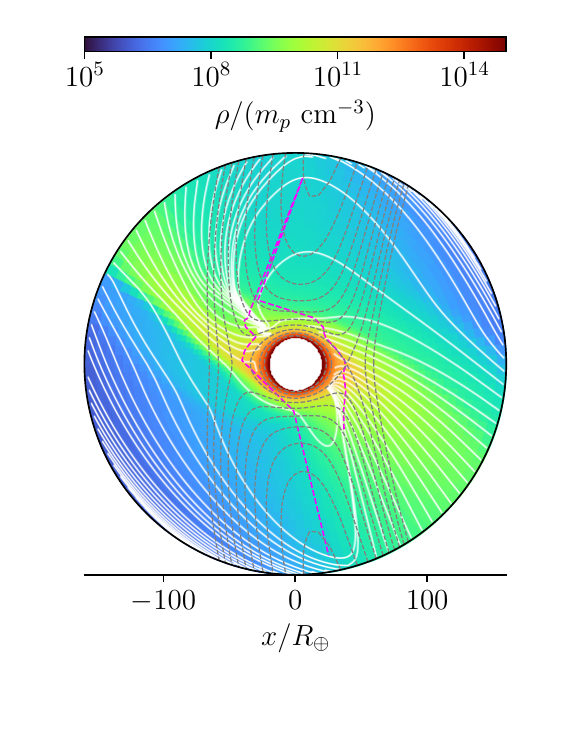}
  \hspace{-0.8cm}
  \includegraphics[width=0.33\linewidth,valign=t]
  {\figdir/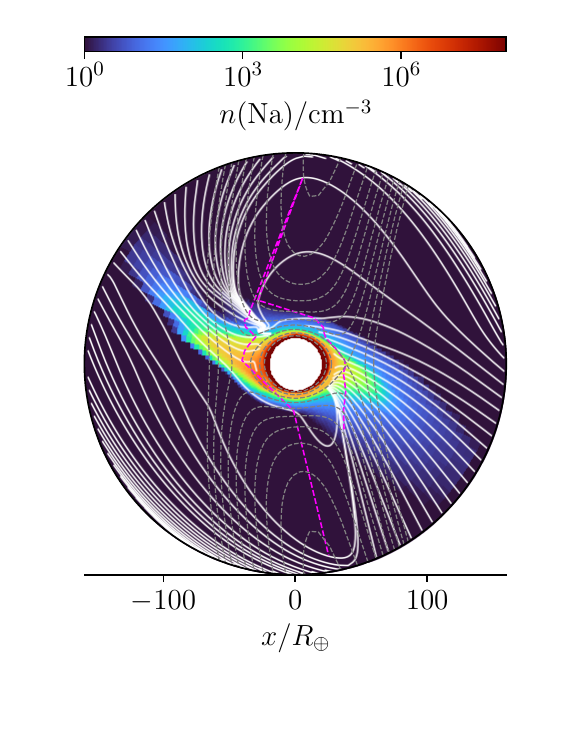}
  \includegraphics[width=0.32\linewidth,valign=t]
  {\figdir/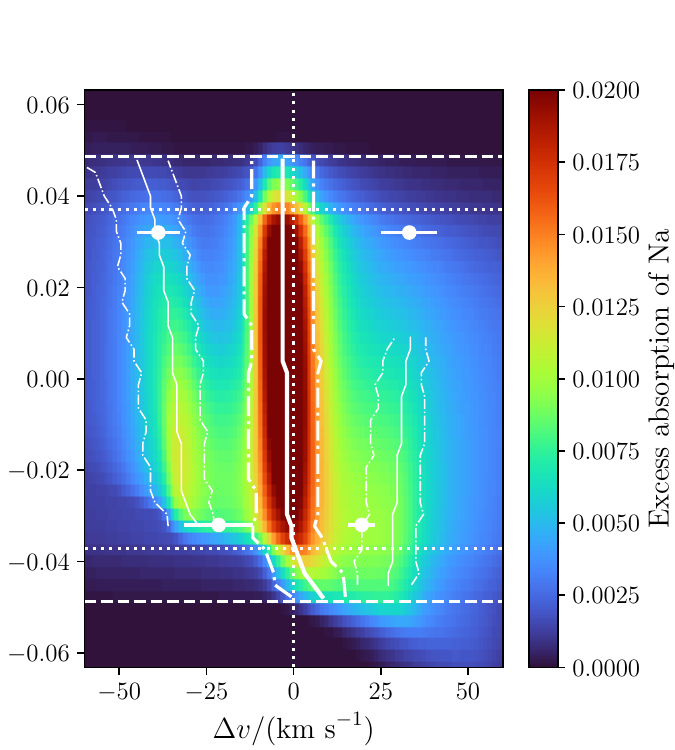}  
  \\ \vspace{-0.8cm}
  \raggedright DUST \\ \vspace{-0.2cm} \hspace{-1.0cm}
  \includegraphics[width=0.33\linewidth,valign=t]
  {\figdir/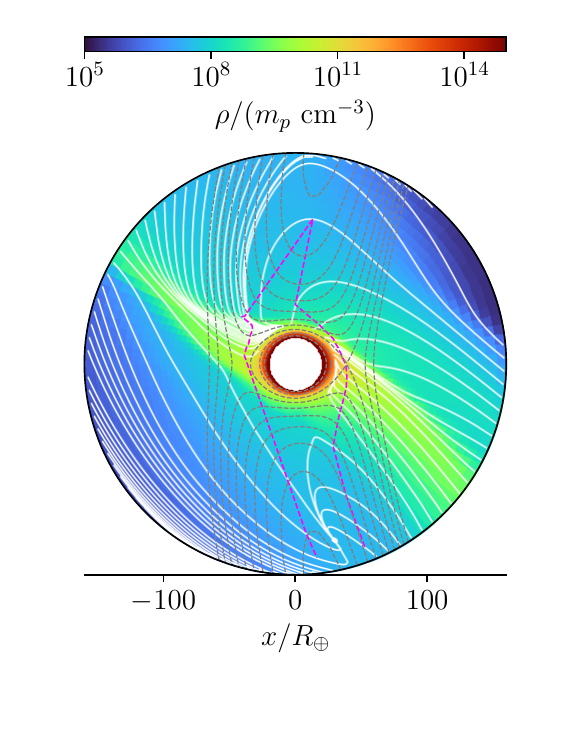}
  \hspace{-0.8cm}
  \includegraphics[width=0.33\linewidth,valign=t]
  {\figdir/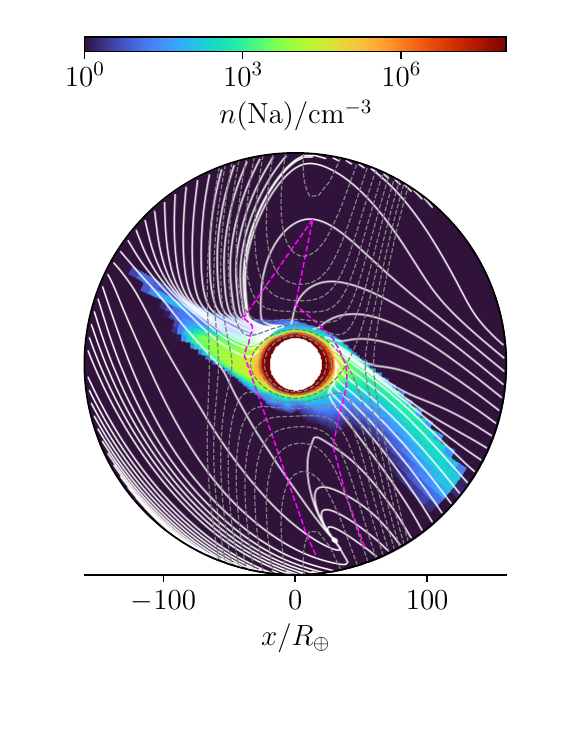}
  \includegraphics[width=0.34\linewidth,valign=t]
  {\figdir/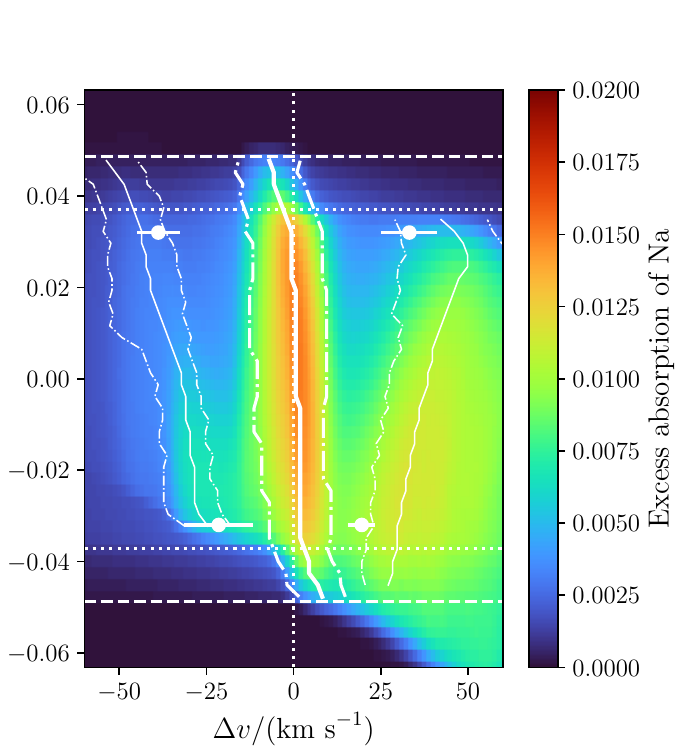}
  \\ \vspace{-0.8cm}
  \raggedright WIND \\ \vspace{-0.2cm} \hspace{-1.0cm}  
  \includegraphics[width=0.33\linewidth,valign=t]
  {\figdir/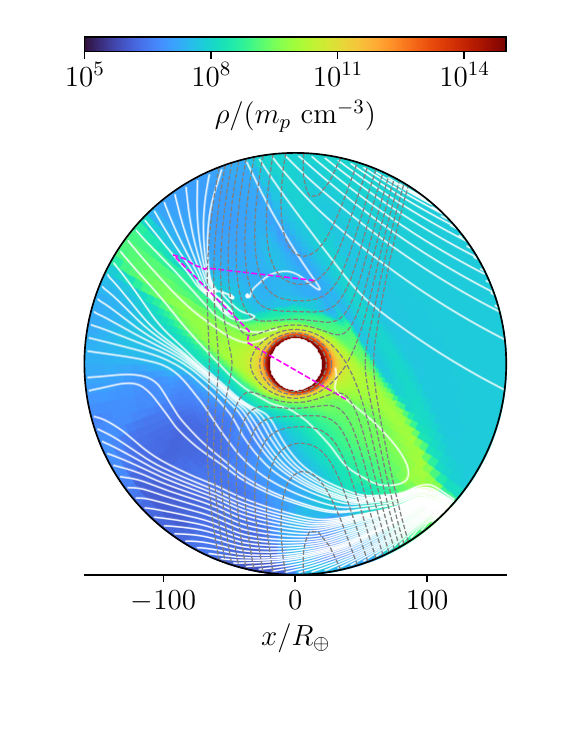}
  \hspace{-0.8cm}
  \includegraphics[width=0.33\linewidth,valign=t]
  {\figdir/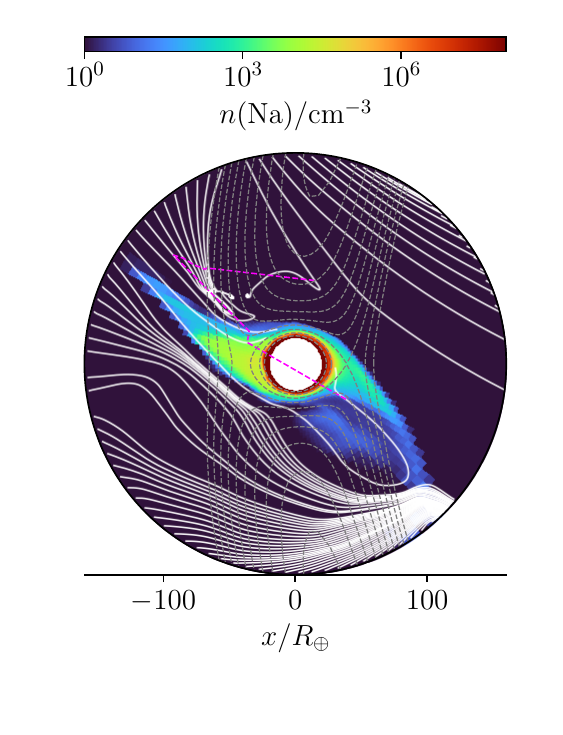}
  \includegraphics[width=0.34\linewidth,valign=t]
  {\figdir/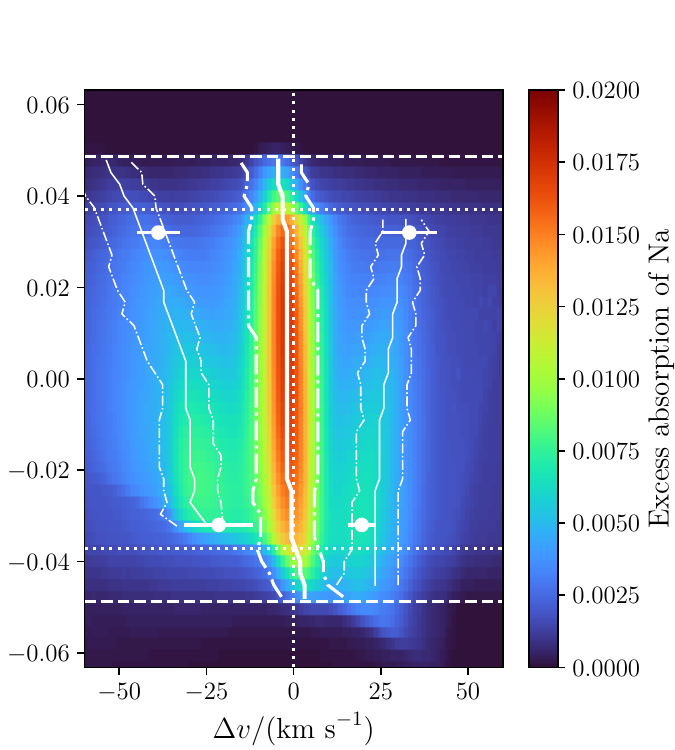}
  \vspace{-0.8cm}
  \caption{Similar to Figure~\ref{fig:multimodel}, for
    models XR10, DUST, and WIND. }
  \label{fig:multimodel-cont}
\end{figure*}

\begin{figure*}
  OPT3 \hspace{0.425\linewidth} FUV10
  \\
  \includegraphics[width=0.48\linewidth]
  {\figdir/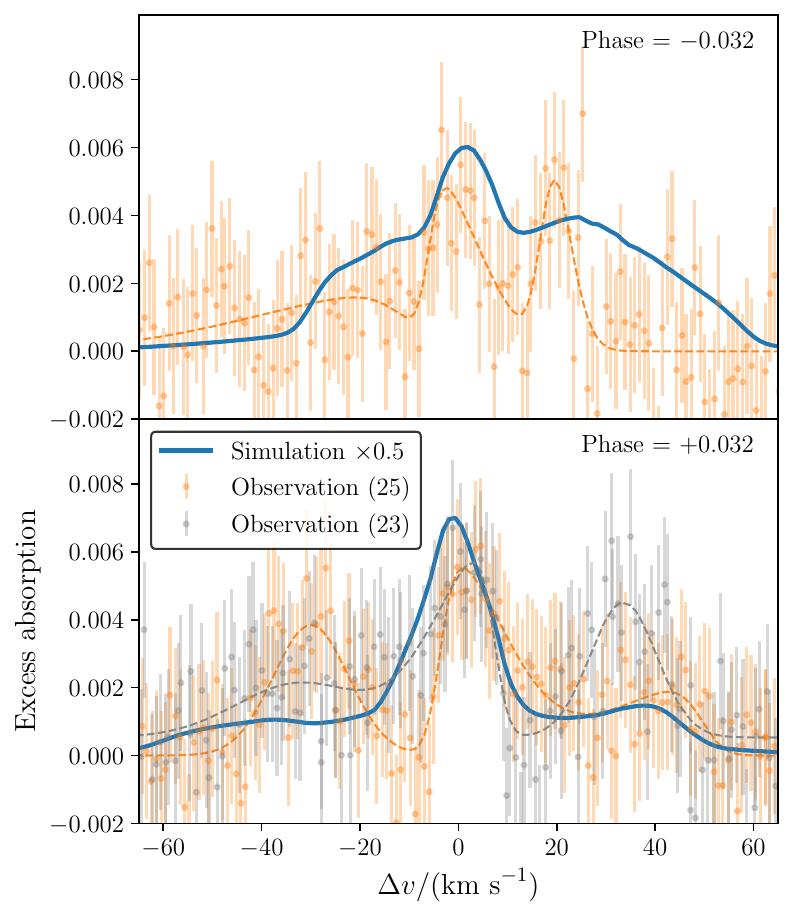}
  \includegraphics[width=0.48\linewidth]
  {\figdir/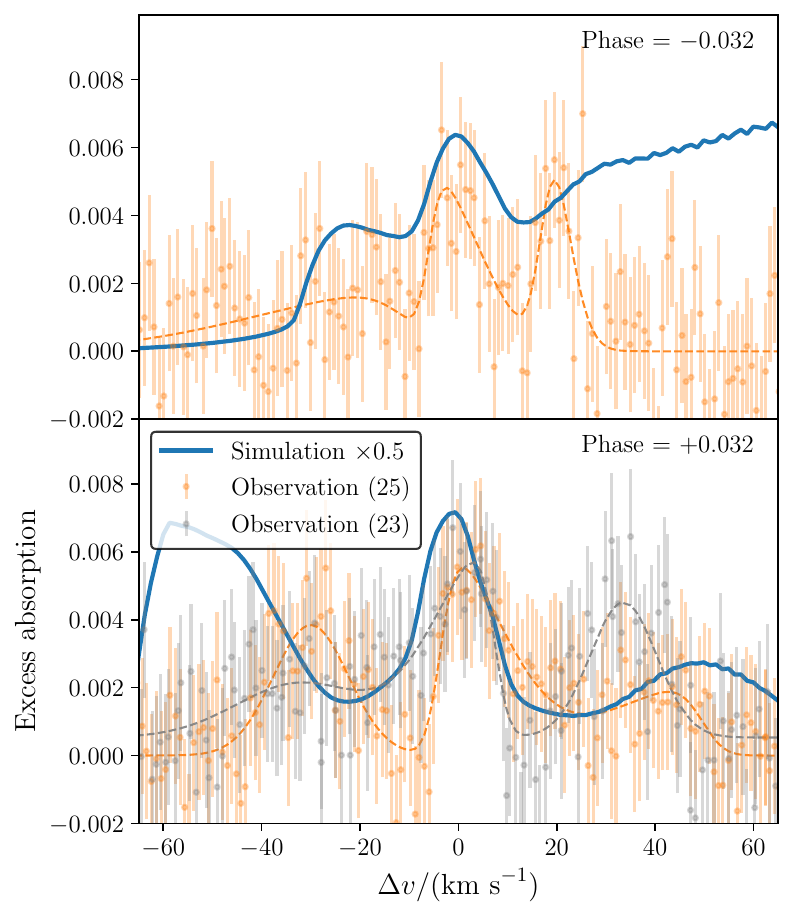}  
  \\
  XR10 \hspace{0.425\linewidth} DUST
  \\
  \includegraphics[width=0.48\linewidth]
  {\figdir/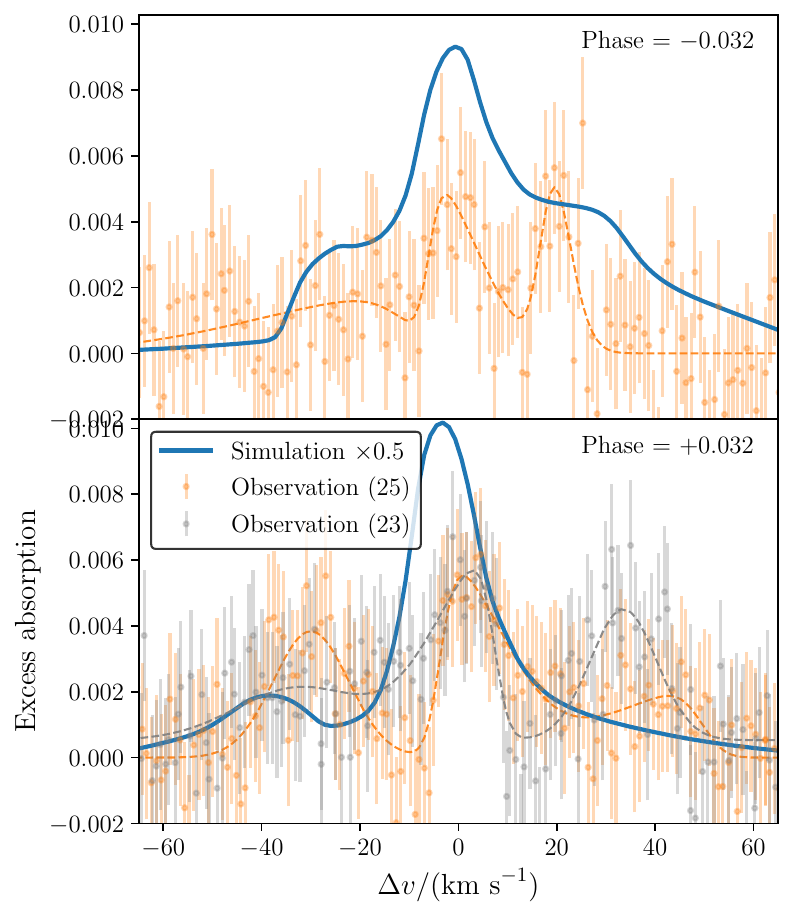}
  \includegraphics[width=0.48\linewidth]
  {\figdir/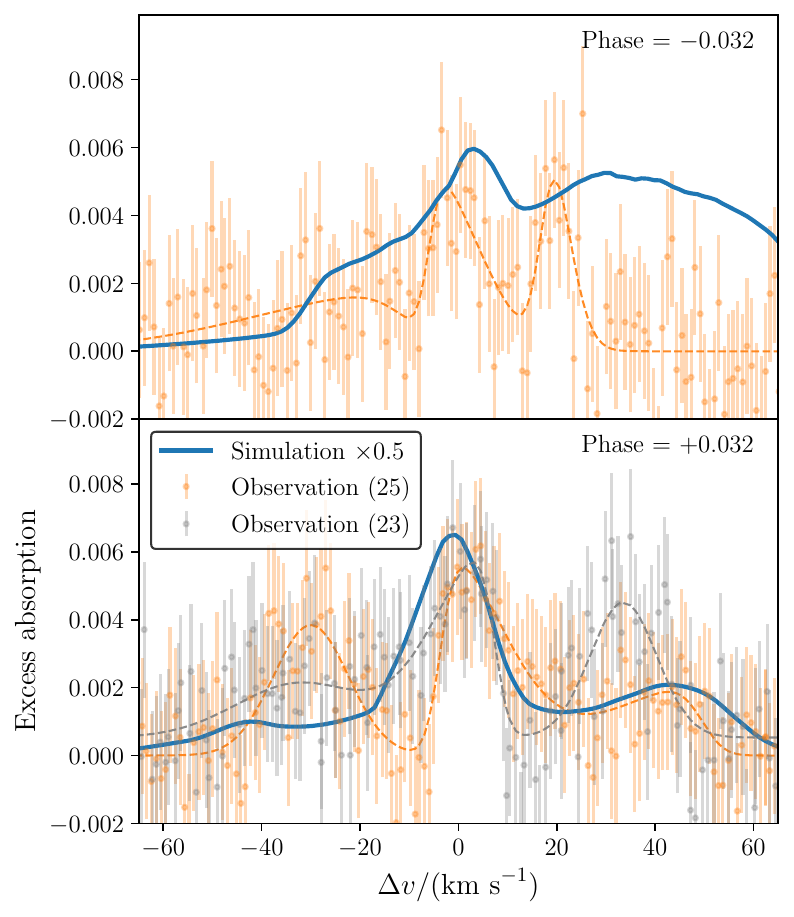}  
  \caption{Similar to Figure~\ref{fig:compare-na-fid}, but
    for Models OPT3 (upper left), FUV10 (upper right),
    XR10 (lower left), and DUST (lower right). }
  \label{fig:compare-na-var}
\end{figure*}

Model XR10, with enhanced X-ray flux, produces Fe, and
H$\alpha$ absorption features very similar to those of the
fiducial Model 0. Notably, the Na absorption feature is more
consistent with observations compared to the fiducial model
(Figures~\ref{fig:multimodel-cont} and
\ref{fig:compare-na-var}).  However, the equivalent width of
the metastable He absorption line is approximately $\sim 5$
times stronger than the observed value. Apart from these
tracers, X-ray photons could potentially drive significant
changes in the deeper atmospheric layers.  The
  penetration depth of X-ray photons ($h\nu = 3~\keV$) is
  estimated as
  $N_{\rm H}(E_{\rm X}) \sim \sigma_{\rm X}^{-1} \sim
  10^{25}~{\rm cm}^{-2}$ for the soft X-ray absorption cross
  section in a hydrogen-dominated atmosphere, corresponding
  to column depths very close to the inner boundary of our
  computational domain (see Figure~\ref{fig:slice-ncol-h}).
  While this estimates indicate that the current resolution
  is marginally sufficient to capture the bulk energy
  deposition of soft X-rays in the simulated layers, the
  detailed photochemistry and thermal structure depend on
  the precise photon energy distribution and secondary
  ionization cascades, which we leave for future
  investigations.  Model EUV10, with enhanced extreme
ultraviolet flux, launches a much broader set of ``extended
spiral arms'' on top of the depleted dense spiral arms,
making the spiral structures more diffuse and
extended. Meanwhile, the neutral species in these arms are
effectively destroyed by the unattenuated FUV and EUV
radiation. Because the recombination rate of ions like
\chem{Na^+} scales roughly with the square of the gas
density, the low-density extended spiral arms cannot
maintain a sufficient population of neutral Na atoms,
leading to a suppression of both the spatial and velocity
extent of the Na absorption features. At the same time, the
stronger EUV flux significantly enhances metastable helium
absorption by ionizing more helium, which then recombines
more efficiently, increasing the equivalent width of the He
$10830~\ang$ line by an order of magnitude relative to the
observed value.

\subsection{Dependence on Stellar Wind Properties}
\label{sec:results-var-wind}

The fiducial Model 0 assumes no impinging stellar wind. A
test run with a weak stellar wind (mass-loss rate
$\dot{M}_{\rm wind} = 10^{-14}~M_{\odot}~\yr^{-1}$; not
shown in this paper) also yields observational results
nearly identical to the fiducial case. This consistency
aligns with the observed persistence of \ext{He} absorption
over extended time periods, as a stronger wind would
otherwise disrupt or remove the extended spiral arms from
the planet's vicinity, thereby spatially truncating the He
$10830~\ang$ absorption features.

Although unlikely to exist for such a F6V star as WASP-121,
a much stronger stellar wind could interact dynamically with
the planetary outflow via a contact discontinuity, exerting
a ram pressure that can confine, compress, and reshape the
escaping atmosphere. Model WIND explores this scenario with
a substantially enhanced stellar wind (mass-loss rate
$\dot{M}_{\rm wind} = 6 \times
10^{-12}~M_{\odot}~\mathrm{yr}^{-1}$ at $400~\km~\s^{-1}$
radial speed, plus a $200~\km~\s^{-1}$ orbital motion
component). The increased ram pressure effectively confines
the planetary outflow on the dayside, compressing it into a
narrower structure, while leaving the nightside outflow and
spiral arms relatively intact
(Figure~\ref{fig:multimodel-cont}, row WIND). This dayside
confinement yields generally similar velocity structures in
the excess absorption features of Fe, Na, and H$\alpha$
compared to the fiducial model.  The amplitude of the Na
absorption during ingress becomes slightly stronger due to
the higher gas density in the compressed region, which
enhances recombination rates and thus maintains a higher
neutral sodium fraction against FUV photoionization.

Notably, the equivalent width of the He $10830~\ang$
absorption line at mid-transit is drastically enhanced in
the Model WIND. The wind-induced compression increases the
density at the interface between the dense spiral arm and
the more extended outflow, thereby boosting the rate of
\chem{He^+} recombination into the metastable triplet state
\ext{He}. Consequently, metastable helium absorption serves
not only as a tracer of the escaping atmosphere but also as
a sensitive indicator of wind-planet interactions, analogous
to the diagnostic role proposed in earlier studies of
evaporating exoplanets \citep{2021ApJ...914...98W,
  2021ApJ...914...99W}.

\section{Discussion and Summary}
\label{sec:discussion}

This study presents a new series of three-dimensional
hydrodynamic simulations of the ultra-hot Jupiter WASP-121b,
which are self-consistently co-evolved with ray-tracing
radiative transfer and non-equilibrium thermochemistry. By
utilizing the GPU-accelerated \kratos{} framework, our
simulations span from the inner atmospheric circulation
regions near the planetary surface out to the extended,
escaping exosphere, thereby overcoming the limitations of
altitude ranges inherent in most GCM simulations.

The fiducial simulation reveals that the inner atmospheric
kinematics are characterized by day-to-night circulation
patterns modulated by the planet's tidally locked spin,
featuring prominent anticyclonic flow on the dayside and
cyclonic structures on the nightside. At higher altitudes,
the atmospheric structure is dominated by a supersonic
photoevaporative outflow, which is shaped by the combined
effects of orbital motion, stellar gravity, and the Coriolis
force into two distinct, relatively dense spiral arms with
gas densities $\rho \gtrsim 10^{10}~m_p~\cm^{-3}$. These
dense arms serve as the foundation for more extended,
attenuated structures ($\rho \sim 10^{8}~m_p~\cm^{-3}$) that
are ionized and heated by EUV irradiation.

Different chemical species act as tracers for distinct
regions within this outflowing atmosphere. Neutral Fe
primarily probes the inner, rotation-dominated layers
influenced by planetary spin and day-night circulation. The
H$\alpha$ and He $10830~\ang$ absorptions trace the
interface between dense and attenuated gas in the spiral
arms, where recombination of \chem{H^+} populates the $2s$
state of neutral hydrogen. Neutral Na, which survives in the
dense spiral arms due to efficient recombination balancing
FUV photoionization, exhibits significant velocity shifts
(approximately $20~\km~\s^{-1}$ redshift during ingress and
$35~\km~\s^{-1}$ blueshift during egress). These shifts,
which are comparable to or exceed the planetary escape
velocity, indicate that the large-scale spiral arm geometry
and kinematics provide a more consistent explanation for the
observed spectral line profiles than local super-rotation or
other surface-bound phenomena.

The parametric study reveals the complex interactions
between stellar radiation, atmospheric escape, and
observable tracers. Enhanced FUV radiation drives more
vigorous outflows but simultaneously photoionizes neutral
sodium and metastable helium, leading to a non-linear
response in absorption line strengths. Stronger EUV flux
expands the spiral arm structures while effectively
destroying neutral species, suppressing sodium features but
significantly enhancing He $10830~\ang$ absorption through
increased recombination to the metastable state. Optical
radiation primarily controls the outflow mass-loss rate by
heating up the gas, while the inclusion of refractory grains
as opacity sources also enhances atmospheric heating and
escape. Strong stellar wind confinement compresses the
dayside outflow, increasing local densities and
recombination rates, which in turn strengthens both sodium
and metastable helium absorption signatures. This
demonstrates that metastable helium serves not only as an
atmospheric escape tracer but also as a sensitive probe of
wind-planet interactions.


\subsection{Comparison with Prior Models}

The 3D non-equilibrium simulations here complement the
growing body of hydrodynamic escape models for WASP-121b.
The thermochemical network shares its foundation with the
established models of \citet{2013Icar..226.1678K} and
\citet{2011ApJ...737...15M} for irradiated hot Jupiter
atmospheres, with extensions to higher-energy photochemistry
as in \citet{2017ApJ...851..150H} (detailed comparisons are
presented in Appendix~\ref{sec:appendix-reactions}). In
addition, our network also includes metal tracers relevant
for transmission spectroscopy diagnostics (Mg, Ca, Fe, Na)
and updates several rate coefficients (dielectronic
recombination, charge exchange) to values appropriate for
the $\sim 10^4$~K thermospheric environment. 

In terms of geometry and dynamics,
\response{\citet{2023ApJ...951..123H} computed
  one-dimensional mass-loss rates for WASP-121b using
  time-dependent hydrodynamic escape models that include
  atomic metal species (Mg, Fe, Ca) and couple the
  hydrodynamics to a full \lya Monte-Carlo radiative
  transfer calculation of the excited hydrogen
  population. Their total mass-loss rate
  ($\sim 2\times 10^{11}~\g~\s^{-1}$ under moderate EUV
  conditions) is about two orders of magnitude lower than
  our fiducial 3D simulation
  ($1.2\times 10^{13}~\g~\s^{-1}$, see also
  Table~\ref{table:models-var}). This difference likely
  reflects the efficient Roche-lobe overflow captured by the
  3D simulations, which allows the lower-energy FUV photons
  to dominate the mass loss (Model FUV10 in
  \S\ref{sec:results-var-rad}), whereas such photons are
  not sufficiently energetic to launch efficient outflows
  under spherical symmetry. A direct quantitative
  comparison is further complicated by differences in the
  treatment of radiative transfer and the lower boundary
  condition. In either case, consistent thermochemistry is
  necessary to obtain adequate planetary mass-loss rates in
  escape simulations \citep{2018ApJ...860..175W}.} The
qualitative distinction is also observed, that the 1D
geometry cannot reproduce the poloidal and azimuthal
structures, especially the spiral arms, while our
simulations identify as essential for the phase-dependent
velocity patterns in Na and H$\alpha$. The dynamic outflows
portrayed in 3D, with timescales comparable to some key
thermochemical processes (e.g., the $\sim 2.2~{\rm hr}$
decay timescale of \ext{He}), in turn necessitates the
inclusion of non-equilibrium thermochemistry in the
simulations.

\subsection{Model Limitations}

The present simulations contain several approximations that
bound their interpretative scope, and we discuss them in
turn.  The fiducial simulation uses a relatively coarse
resolution
($N_r \times N_\theta \times N_\phi = 128 \times 32 \times
128$) focusing on the stability of the densest regions,
which resolves the bulk spiral-arm morphology but likely
under-resolves small-scale mixing at the interface between
dense and attenuated arms.  The computational domain is
restricted to the northern hemisphere, enforced by an
equatorial symmetric boundary condition. Nonetheless,
convergence tests varying the resolution and domain size
(not directly shown here), as well as dedicated re-run
experiments that incorporate the \lya post-processing
corrections and the He$^{2+}$ network extensions discussed
in Appendices~\ref{sec:appendix-ha-old}
and~\ref{sec:appendix-hepp}, have confirmed that the
spiral-arm geometry and the transmission spectral profiles
are robust to these numerical parameters.  The inner
boundary is a reflecting surface at $p_{\rm in}=0.83$~bar,
which is not sufficient for deep convective forcing and may
affect the circulation pattern at the lowest modeled
altitudes.  Extending the simulation to deeper pressures
would require a fully coupled treatment of radiative
transfer and convection that is beyond the scope of this
work.

The simulations are run to a quasi-steady state;
time-dependent phenomena such as the Kelvin-Helmholtz
instability (KHI) identified in \S\ref{sec:results} are
captured only to the extent they saturate in the
steady-state solution.  Fully time-dependent simulations
would be needed to study the onset, growth, and decay of
such instabilities and their observational signatures in
time-resolved transmission spectroscopy attempting to
resolve the responses of planetary atmospheres on the
stellar activities. 

One constraint is that the present calculations are purely
hydrodynamic and do not include magnetic fields.  In the
partially ionized thermosphere of UHJs, magnetic effects may
compete with thermal pressure in shaping the outflow
morphology, particularly at the L1 and L2 Lagrangian points,
where the simulation yields number densities
$n \sim 10^{11}$--$10^{12}~m_p~\cm^{-3}$ and temperatures
$T \sim 10^4~\K$.  The plasma beta,
$\beta = 8\pi n k_B T / B^2$, then falls in the range
$\beta \sim 0.4-4$ for $B \sim 3~{\rm G}$--in other words,
it straddles unity and is generally sub-thermal with
sufficiently strong magnetic fields. Nonetheless, the
magnetic fields of WASP-121b still cannot be directly
measured. The only available constraint,
$B \lesssim 3~{\rm G}$, is an indirect inference from
model-dependent atmospheric drag signatures in the NIRISS
phase curve \citep{2026ApJ..1004..102F}.  We therefore
acknowledge the absence of magnetohydrodynamics (MHD) as a
prospective caveat, as the spiral-arm morphology presented
here could be modified by magnetic confinement if the actual
field is comparable to the thermal energy density in the
outflow, a regime that is not ruled out by the available
observations. Another limitation is that the radiation field
is treated by ray-tracing in the irradiation band and by
simplified recipes (local escape probability;
\S\ref{sec:method-chem}) for line cooling in the IR and
optical bands.  A multi-band radiative transfer treatment
within MHD frameworks, although computationally more
expensive, would provide a more rigorous cooling computation
and should be pursued as computational resources permit.  

\subsection{Future works}

The synergy between advanced numerical modeling and
state-of-the-art observational instrumentation are promising
to deepen our understanding of UHJs and planetary
atmospheres in general. For instance, the extended \ext{He}
trails observed by JWST for WASP-121b, which stretch well
beyond the transit window \citep{2025NatCo..1610822A}, can
be self-consistently modeled within the \kratos{} framework
by extending the simulation domain (e.g., to
$r=240~R_\oplus$ from the planet surface) and leveraging
increased GPU computational resources. Simultaneous
multi-wavelength observations of species that trace
different atmospheric layers, such as He (from JWST) and Fe,
H$\alpha$, and Na (from ground-based facilities), will allow
us to probe potential variability and chemical
stratification in a coherent manner. This can be achieved by
combining high-resolution spectrographs like KPF
\citep{Gibson2016}, HISPEC \citep{Konopacky2023} at the Keck
Observatory, and PEPSI \citep{Strassmeier2015} and iLocater
\citep{Crass22} at the Large Binocular Telescope.

Time-resolved spectral monitoring during individual transits
can reveal dynamical variability, such as the changing
velocity patterns of neutral Fe lines reported by
\citet{Ehrenreich_2020, PIRANGA2, PIRANGA3}. Within the
\kratos{} framework, such variability can be interpreted in
terms of deeper atmospheric heating processes, potentially
driven by opacity sources like carbon grains. Looking ahead,
future extremely large telescopes (ELTs) will provide the
spectral resolution, time resolution, and SNR that are
necessary to detect short-term variability induced by
hydrodynamic instabilities such as the Kelvin-Helmholtz
instability (KHI) discussed in \S\ref{sec:results}, offering
insights into the real-time dynamics of exoplanetary upper
atmospheres.  Further inclusion of magnetic fields and
non-ideal magnetohydrodynamic effects (whose magnetic
diffusivity parameters are determined consistently by
thermochemical calculations) could also open a new window
for understanding the interplay between planetary outflows
and stellar properties, and prospectively bridging the gap
between global circulation models and microphysical escape
processes.

\begin{acknowledgments}
  This work is supported by the National Natural Science
  Foundation of China (NSFC) under Grant 12573067. The
  computational resources supporting this work are provided
  by the Kavli Institute for Astronomy and Astrophysics,
  Peking University. We thank our colleagues: Fei Yan, Meng
  Sun, Wei Wang, Bin Ren, Jun Yang, Siyi Feng, for helpful
  discussions and suggestions on the contents of the paper.
\end{acknowledgments}



\bibliography{uhj}
\bibliographystyle{aasjournal}

\clearpage
\appendix

\begin{deluxetable}{l c c c c c}
\tablecaption{Species coverage across network generations}
\label{table:appendix-compare}
\tablehead{
  \colhead{Species / class} &
  \colhead{This work} &
  \colhead{M11} &
  \colhead{K13} &
  \colhead{H17} &
  \colhead{H23}
}
\startdata
\multicolumn{6}{c}{Hydrogen and Helium} \\
H, H$_2$, H$^+$, H$_2^+$, H$_3^+$ &
  5 & 5 & 5 & 5 & 2$^e$ \\
H$^{-}$ &
  \checkmark{} & -- & -- & -- & -- \\
He, He$^+$ &
  2 & 1$^a$ & 2 & 1$^a$ & -- \\
He$^\ast$(2$^3$S) &
  \checkmark & -- & -- & --$^b$ & -- \\
\hline
\multicolumn{6}{c}{ C--O--H chemistry} \\
C, C$^+$, CH, CH$^+$, CO &
  5 & 5$^c$ & 5 & 5 & -- \\
O, O$^+$, O$_2$ &
  3 & 3$^d$ & 3 & 3 & -- \\
OH, OH$^+$, H$_2$O, H$_2$O$^+$, H$_3$O$^+$ &
  5 & 5 & 5 & 5 & -- \\
\hline
\multicolumn{6}{c}{Metal and ions} \\
Na, Na$^+$ &
  \checkmark & -- & -- & \checkmark & -- \\
Fe, Fe$^{+}$ &
  \checkmark & -- & -- & -- & \checkmark \\
Mg, Mg$^{+}$ &
  \checkmark & -- & -- & -- & \checkmark \\
Ca, Ca$^{+}$ &
  \checkmark & -- & -- & -- & \checkmark \\  
\hline
\enddata \tablecomments{\response{Entries report the number
    of species within each class tracked by the
    corresponding network where a number is given, and a
    checkmark or dash indicates the presence or absence of
    that specific species where no count applies.} The
  reaction network is constructed based on the same network
  in \citet{2021ApJ...914...98W, 2021ApJ...914...99W}, in
  which He$^\ast$(2$^3$S) introduced following
  \citet{2018ApJ...855L..11O}.  M11 includes a wider C--O--N
  network ($\sim$300 reactions) with nitrogen-bearing
  species.  K13 models the thermosphere with $\sim$100
  reactions.  Details in the main text
  (\S\ref{sec:discussion}). \\
  $^a$ He only, no He$^+$. \\
  $^b$ H17 tracks $\chem{H}^{2s}$ as the excited state for
  H$\alpha$, not He$^\ast$(2$^3$S). \\
  $^c$ M11 additionally tracks N-bearing hydrocarbons
  and C$_2$ species beyond the common C--O set. \\
  $^d$ M11 includes O($^1$D) and nitrogen oxides beyond the
  ground-state O--O$_2$ chemistry. \\
  \response{
  $^e$ H23 tracks H and H$^+$ (with the excited hydrogen
  states for the H$\alpha$ diagnostics) and the atomic metal
  species Mg, Fe, and Ca \citep{2023ApJ...951..123H}.
}}
\end{deluxetable}

\begin{deluxetable}{l l l}
\tablecaption{Photochemical reactions}
\label{table:appendix-photo}
\tablehead{ \colhead{Reactant} & \colhead{Product(s)} &
  \colhead{Shielding }}
\startdata
H    & H$^{+}$ + e$^{-}$ & -- \\
H$_2$ & H + H & H$_2$ \\
& H$^{+}$ + H$^{-}$ & -- \\
& H$^{+}$ + H + e$^{-}$ & -- \\
& H$_2^{+}$ + e$^{-}$ & -- \\
H$^{-}$ & H + e$^{-}$ & -- \\
H$_2^{+}$ & H + H$^{+}$ & -- \\
H$_2$O$^{+}$ & H + OH$^{+}$ & -- \\
H$_3^{+}$ & H + H$_2^{+}$ & -- \\
& H$^{+}$ + H$_2$ & -- \\
He   & He$^{+}$ + e$^{-}$ & -- \\
He$^{+}$ & He$^{2+}$ + e$^{-}$ & -- \\
He$^\ast$ & He$^{+}$ + e$^{-}$ & -- \\
C    & C$^{+}$ + e$^{-}$ & C, H$_2$ \\
CO   & C + O & CO, C, H$_2$ \\
CH   & C + H & -- \\
& CH$^{+}$ + e$^{-}$ & -- \\
CH$^{+}$ & C$^{+}$ + H & -- \\
& C + H$^{+}$ & -- \\
O    & O$^{+}$ + e$^{-}$ & -- \\
O$_2$ & O + O & -- \\
OH   & O + H & -- \\
& OH$^{+}$ + e$^{-}$ & -- \\
OH$^{+}$ & O$^{+}$ + H & -- \\
H$_2$O & OH + H & -- \\ 
& H$_2$O$^{+}$ + e$^{-}$ & -- \\
Na   & Na$^{+}$ + e$^{-}$ & -- \\
Mg   & Mg$^{+}$ + e$^{-}$ & -- \\
Fe   & Fe$^{+}$ + e$^{-}$ & -- \\
\enddata
\tablecomments{Species for which self-sheilding and
  cross-shielding are computed using fitted column-density
  attenuation factors. See also \citet{1996ApJ...465..487V},
  \citet{2023AAP...675A..25H}, and references therein. }
\end{deluxetable}

\section{Thermochemical Network}
\label{sec:appendix-reactions}

The chemical network comprises 33 gas-phase species: 14
neutrals, 16 cations, 1 anion, plus the electron (negatively
charged), the He$^\ast$(2$^3$S) metastable helium state, and
one dust grain species (Gr) that is only involved in the
DUST test run (\S\ref{sec:results-var-rad}). This set of
species tracks across 147 thermal reactions, 29
photochemical reactions, 4 molecular cooling processes, and
5 atomic cooling processes.  All thermal rate coefficients
are sourced from the UMIST 2012 database for astrochemistry
\citep{2013A&A...550A..36M}. \response{The 5 metal atomic
  cooling processes are the fine-structure lines of C$^+$
  (158~$\mu$m), C (370 and 609~$\mu$m), and O (63 and
  146~$\mu$m) based on the data compiled in Chapter~3 of
  \citet{2011piim.book.....D} (and references therein),
  treated via the escape-probability approach
  (\S\ref{sec:method-chem}). \lya cooling is not included in
  the fiducial model. As stated in Chapter~3 of
  \citet{2011piim.book.....D} and confirmed by Appendix~A of
  \citet{2018ApJ...860..175W}, \lya cooling alone is
  generally not sufficient to establish a steady state in
  photoevaporating planetary atmospheres, and the metal
  fine-structure lines could be the more important coolants
  at a few times $\sim 10^3$~K temperatures near the bases
  of outflows.  To verify this choice, we performed an
  additional test run with \lya cooling enabled (6 atomic
  cooling processes in total) and otherwise identical
  setups. All diagnostics, including the slice plots, the
  spatial distributions of the hydrodynamic and
  thermochemical quantities, and the synthetic spectra, are
  very similar to those of the fiducial model; the mass-loss
  rate is reduced to $0.62\times 10^{-7}~M_\oplus~\yr^{-1}$
  (Table~\ref{table:models-var}), a change of less than
  5\%. We nonetheless note that the escape-probability
  approach for \lya may be subject to non-negligible
  uncertainties, and fully self-consistent radiative
  transfer simulations would be desirable in future work.  }

\begin{figure}
  \centering
  \includegraphics[width=0.8\linewidth]
  {\figdir/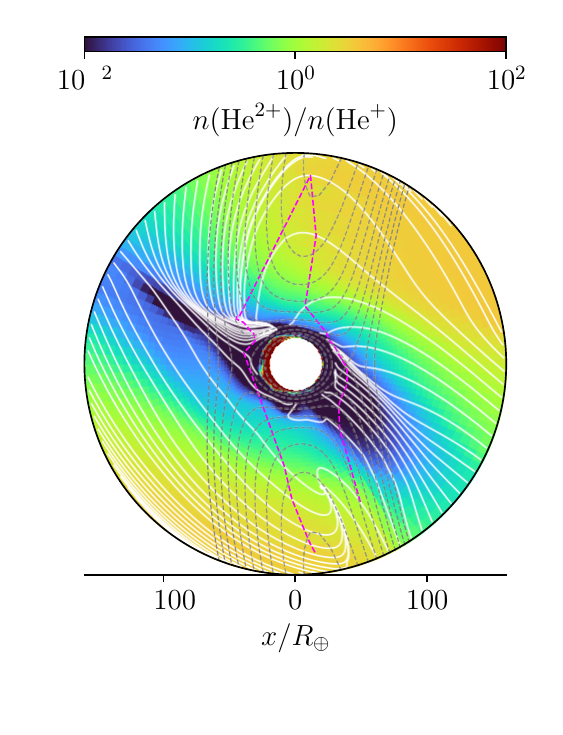}  
  \caption{Slice plot in the equitorial plane (similar to
    Figure~\ref{fig:slice-fid}), showing the ratio of
    particle number density of $\chem{He}^{2+}$ to
    \chem{He^+}. }
  \label{fig:appendix-hepp}
\end{figure}


 

As a full presentation of the gas-phase reactions
involved could be tedious and redundant, we summarize the
species involved as a key inspection into the network
structure and compare its coverage with that of
representative published networks for irradiated exoplanet
atmospheres.  Table~\ref{table:appendix-compare} maps the
species tracked in our network against those included in the
established networks of \citet{2011ApJ...737...15M} (Moses
et al., M11), \citet{2013Icar..226.1678K} (Koskinen et al.,
K13), and \citet{2017ApJ...851..150H} (Huang et al., H17).
A ``\checkmark{}'' denotes presence; dashes indicate absence
or an alternative treatment.


Table~\ref{table:appendix-photo} specifically lists the 29
photochemical reactions in the network. Data necessary for
calculating photoionization and photodissociation
cross-section are taken from \citet{1996ApJ...465..487V} and
the standard \citet{2006agna.book.....O} compilation,
integrated respectively over the eight radiation bands
listed in Table~\ref{table:fiducial-pars}.  The
self-shielding and cross-shielding effects are also taken
into account by ray-tracing column density calculations
along with the parameterizations within
\citet{2011piim.book.....D} for molecular hydrogen
(\chem{H_2}), and \citet{2023AAP...675A..25H} for other
species.  The six metal photoionization channels (Na, Mg,
Fe) and the He$^\ast$(2$^3$S) photoionization channel are
additions that distinguish the UHJ network.  These provide
the radiative pathways needed for tracking UHJ
transmission-spectroscopic diagnostics.

\section{Highly Ionized Helium: Abundance Distribution and
  Impact on Extinction}
\label{sec:appendix-hepp}

A dedicated test simulation was run with the fiducial
radiation and wind parameters, augmented to include explicit
He$^{2+}$ species together with the associated He$^+$
photoionization (He$^+ + h\nu \to$ He$^{2+} + e^-$) and
He$^{2+}$ recombination (He$^{2+} + e^- \to$ He$^+$)
reactions.  The test confirms that He$^{2+}$ quickly becomes
the dominant helium ionization state in the diffuse outer
regions of the simulation domain, where the gas number
density falls below $n \sim 10^8~m_p~\cm^{-3}$.  However,
those regions only contribute a small fraction to the He
$10830~\ang$ optical depth. The metastable helium
population, He($2^3S$), is determined by recombination at
higher densities within the dense spiral arms. Therefore,
the resulting change in the equivalent width of the
He~10830~\ang\ absorption is below $10\%$.
Figure~\ref{fig:appendix-hepp} maps the ratio
$n({\rm He}^{2+})/n({\rm He}^+)$ on the orbital plane at
mid-transit.  Figure~\ref{fig:appendix-hepp} shows this
ratio explicitly: in the regions where He~10830~\ang\
absorption is significant (cf.\ Figure~1 in the main text),
$n({\rm He}^{2+})/n({\rm He}^+) \sim 10^{-2}$ to $10^{-1}$
-- one to two orders of magnitude below unity.  This
confirms that including He$^{2+}$ changes the He* population
by less than $10\%$, justifying its omission from the
fiducial model.

\begin{figure*}
  \centering
  \vspace{0.5cm} 
  \raisebox{-\height}
  {
  \includegraphics[width=0.4\linewidth]
  {\figdir/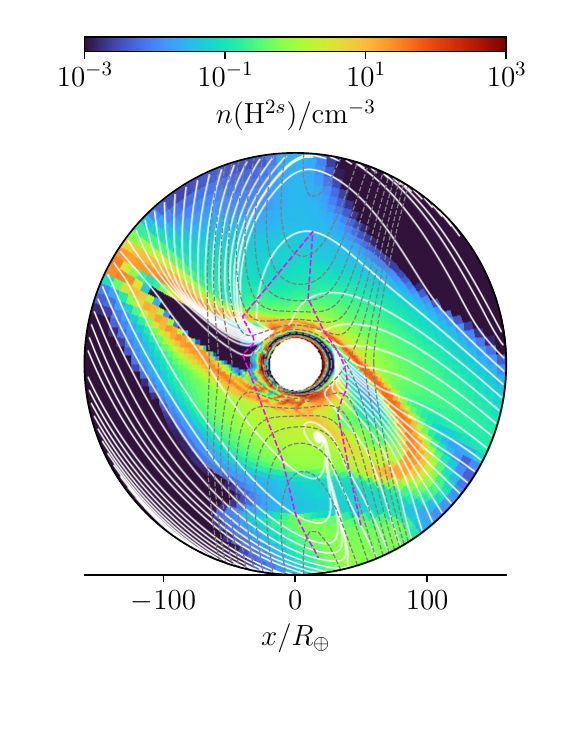}
}
  \hspace{1.8cm}
  \raisebox{-\height}
  {
  \includegraphics[width=0.4\linewidth]
  {\figdir/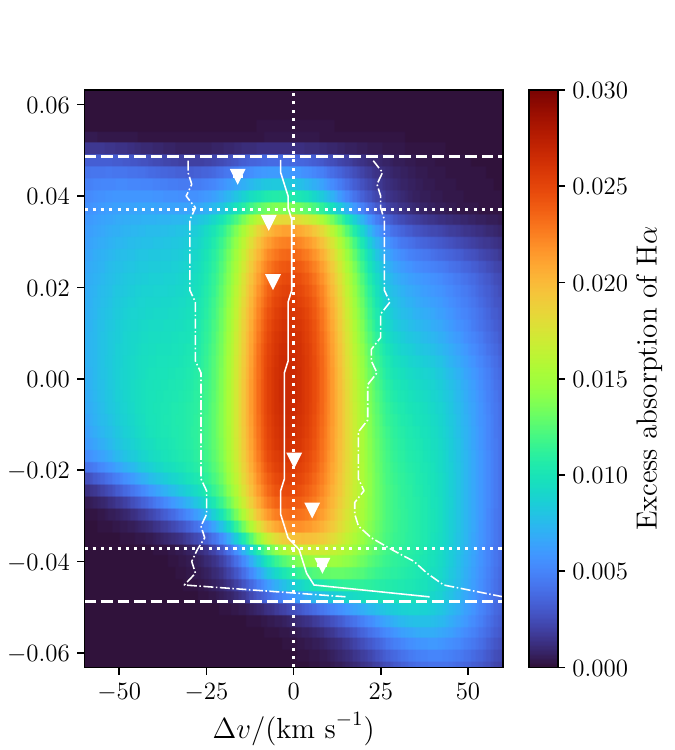}
  }
  \vspace{-1.1cm}
  \caption{Slice plot for $\chem{H}^{2s}$ number density in
    the equitorial plane (left panel; similar to
    Figure~\ref{fig:slice-fid}), and the excess absorption
    spectra by H$\alpha$ (right panel; similar to
    Figure~\ref{fig:spec-fid}), using alternative methods of
    calculating the $\chem{H}^{2s}$ abundance
    (Appendix~\ref{sec:appendix-ha-old}).  }
  \label{fig:appendix-ha-old}
\end{figure*}

\begin{figure*}
  \centering
  \includegraphics[width=0.9\linewidth]
  {\figdir/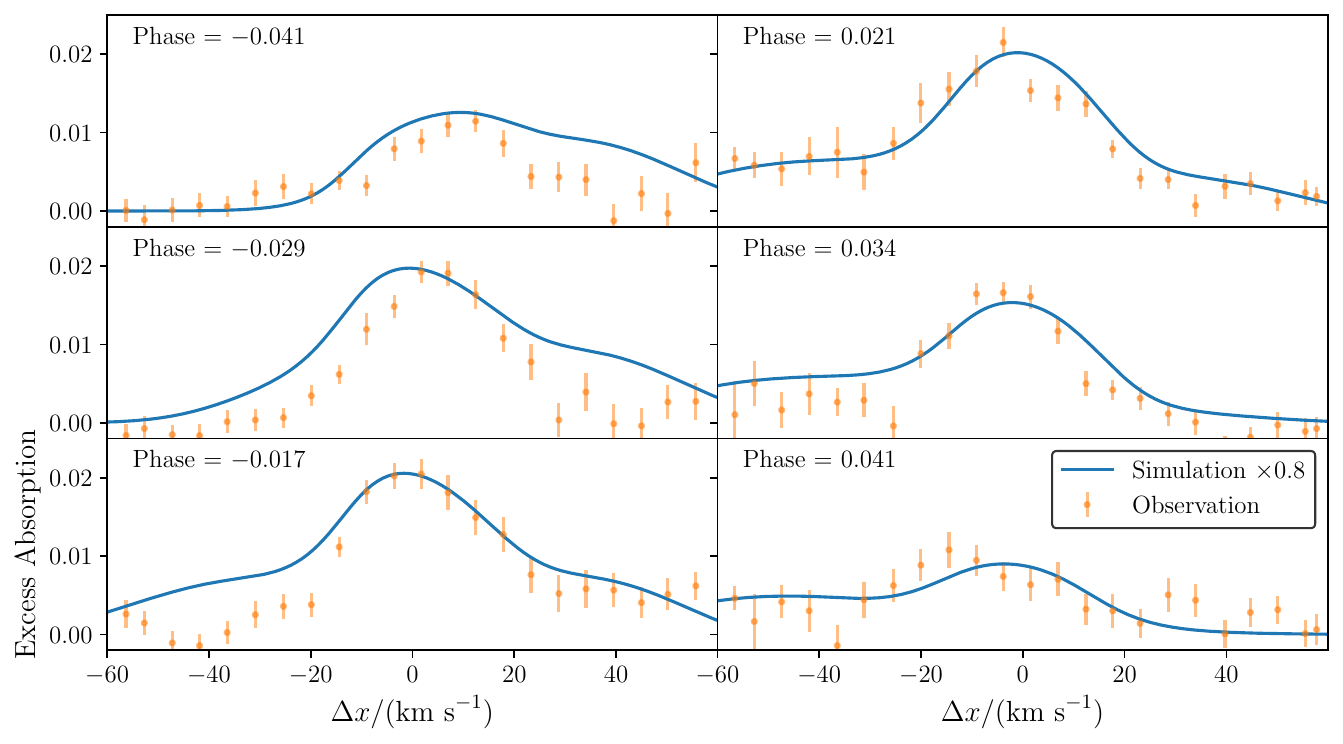}  
  \caption{ Similar to Figure~\ref{fig:compare-halpha-fid},
    but using alternative methods of
    calculating the $\chem{H}^{2s}$ abundance
    (Appendix~\ref{sec:appendix-ha-old}). }
  \label{fig:appendix-compare-ha-old}
\end{figure*}

\section{Alternative H$\alpha$ Population Scheme}
\label{sec:appendix-ha-old}

For transparency, we present the H$\alpha$ transmission
spectra computed using an alternative scheme for the
$\chem{H}^{2s}$ population: no \lya radiative pumping,
with the $\chem{H}^{2s}$ population determined solely by
recombination and charge exchange processes, and the
$2s \to 2p$ collisional level-crossing rate coefficient from
\citet{2013ApJ...772..144C}.
Using this althernative scheme, the spatial redistribution
and time-dependent spectra of the H$\alpha$ absorption are
presented in Figure~\ref{fig:appendix-ha-old}.
The velocity profiles are broadly similar to the fiducial,
as the regions of H$\alpha$ extinction has similar spatial
extents when projected along the lines of sight, because the
\lya-affected regions share a similar velocity
distribution to the recombination-dominated $\chem{H}^{2s}$
regions.  Meanwhile, the amplitudes differ by $\sim 20~\%$
(see also \S\ref{sec:H_alpha}). The \lya pumping
channel, omitted in this alternative $\chem{H}^{2s}$ scheme,
can raise the $\chem{H}^{2s}$ population by up to $\sim 2$
orders of magnitude in the thermospheric regions where
H$\alpha$ absorption forms.  However, the collisional
$2s \leftrightarrow 2p$ level crossing rate coefficient from
\citet{2006agna.book.....O} is $\sim 10^2$ times larger than
the \citet{2013ApJ...772..144C} value used in our that
alternative scheme, and this enhancement suppresses the
$\chem{H}^{2s}$ population by a comparable factor.

\end{CJK*}
\end{document}